\documentclass[useAMS,usenatbib]{mn2e}
\usepackage{amssymb,amsmath,graphicx}
\usepackage{longtable}[1]
\usepackage{graphics}
\usepackage{threeparttable}
\usepackage{lscape}
\usepackage{color}
\usepackage{enumitem}
\usepackage{xspace} 
\usepackage[none]{hyphenat}
\usepackage{enumitem}
\usepackage{blindtext} 
\usepackage[hyphens]{url} 
\usepackage[colorlinks,urlcolor=black, citecolor=blue, linkcolor=black]{hyperref}

\newcommand{\degree}{\ensuremath{^\circ}\xspace}
\newcommand{\Ha}{$\mathrm{H}\alpha$\xspace}
\newcommand{\Hb}{$\mathrm{H}\beta$\xspace}
\newcommand{\NII}{$[\mathrm{N}\textsc{ii}]$\xspace}

\newcommand{\NIIb}{$[\mathrm{N}\textsc{ii}]\,\lambda 6583$\xspace}

\newcommand{\OIIIb}{$[\mathrm{O}\textsc{iii}]\,\lambda 5007$\xspace}
\newcommand{\SII}{$[\mathrm{S}\textsc{ii}]$\xspace}

\hypersetup{draft}

\title[Properties of ionized outflows in MaNGA DR2 galaxies]{Properties of ionized outflows in MaNGA DR2 galaxies}
\author[Rodr\'iguez del Pino et. al.]{Bruno Rodr\'iguez del Pino$^{1}$\thanks{E-mail: brodriguez@cab.inta-csic.es}, 
Santiago Arribas$^{1}$,
Javier Piqueras-L\'opez$^{1}$, 
\newauthor
Montserrat Villar-Mart\'in$^{1}$, 
Luis Colina$^{1}$\\
$^{1}$Centro de Astrobiolog\'ia (CSIC-INTA), Torrej\'on de Ardoz, 28850 Madrid, Spain \\
}

\begin{document}

\date{Accepted ---. Received ---; in original form ---}

\pagerange{\pageref{firstpage}--\pageref{lastpage}} \pubyear{2013}

\maketitle

\label{firstpage}

\begin{abstract}

	We present the results from a systematic search and characterization of ionized outflows in nearby galaxies using the data from the second Data Release of the MaNGA Survey (DR2; $>$ 2700 galaxies, $z\leq0.015$). Using the spatially-resolved spectral information provided by the MANGA data, we have identified $\sim5200$ \Ha-emitting regions across the galaxies and searched for signatures of ionized outflows. We find evidence for ionized outflows in 105 regions from 103~galaxies, roughly 7\% of all the \Ha-emitting galaxies identified in this work. Most of the outflows are nuclear, with only two cases detected in off-nuclear regions. Our analysis allows us to study ionized outflows in individual regions with $SFRs$ down to $\sim0.01$~M$_{\odot}$yr$^{-1}$, extending the ranges probed by previous works. The kinematics of the outflowing gas are strongly linked to the type of ionization mechanism: regions characterized by LIER (Low-Ionization Emission Region) emission host the outflows with more extreme kinematics ($FWHM_{\rm broad}\sim$~900~km/s), followed by those originated in AGN (550~km/s), `Intermediate' (450~km/s) and SF (350~km/s) regions. Moreover, in most of the outflows we find evidence for gas ionized by shocks. We find a trend for higher outflow kinematics towards larger stellar masses of the host galaxies but no significant variation as a function of star formation properties within the SFR regime we probe ($\sim0.01-10$~M$_{\odot}$yr$^{-1}$). Our results also show that the fraction of outflowing gas that can escape from galaxies decreases towards higher dynamical masses, contributing to the preservation of the mass-metallicity relation by regulating the amount of metals in galaxies. Finally, assuming that the extensions of the outflows are significantly larger than the individual star-forming regions, as found in previous works, our results also support the presence of star formation within ionized outflows, as recently reported by \citet{maiolino_star_2017} and  \citet{gallagher_widespread_2018}. 




\end{abstract}


\section{Introduction}

%
%
%

The evolutionary path of galaxies through cosmic time is strongly marked by the availability of gas reservoirs that serve as fuel to perpetuate the formation of new stars and the feeding of their 
central supermassive black holes. However, these gas reservoirs are sensitive to the influence of feedback from supernovae explosions, stellar winds or active galactic nuclei (AGN) that can have a severe impact in the gas, mixing and/or heating it up, or even expelling it from the galaxy. Moreover, the star formation activity can be reduced or halted, preventing the formation of new stars and the subsequent stellar mass growth of galaxies \citep{veilleux_galactic_2005, heckman_implications_2016}. Such relevant consequences that feedback effects can have in the evolution of galaxies demand a proper understanding of their incidence and properties. 

Generally, the study of feedback processes is driven by a two-fold motivation: on the one hand, there is a strong interest in understanding the physics of the feedback phenomenon and its characterization; on the other hand, there is also interest, probably even stronger, in evaluating whether the effects of feedback can help explaining some of the observational results that are still not fully comprehended. For instance, when the effects from supernova, stellar winds and AGN feedback are included in cosmological simulations, the number of galaxies formed at low and high masses is reduced, alleviating the existent discrepancies between the predicted and observed baryonic mass functions \citep{benson_what_2003, benson_early_2003}. Apart from halting and/or enhancing star formation activity, feedback-induced galactic outflows might also play an important role regulating the amount of metals in galaxies \citep{brooks_origin_2007, finlator_origin_2007}, contributing to maintaining the relation between the gas-phase metallicity and stellar mass of the galaxies \citep{tremonti_origin_2004, mannucci_fundamental_2010, chisholm_metal-enriched_2018}. Moreover, the observed metal content of the inter-galactic medium (IGM) in high-redshift galaxies could also be explained by the ejection of inter-stellar material via galactic outflows \citep{aguirre_metal_2001}. Moreover, although feedback processes have been generally invoked as mechanisms that prevent or suppress star formation activity, it has been suggested, first in simulations \citep{ishibashi_active_2012, ishibashi_can_2013} and very recently in observations  \citep{maiolino_star_2017, gallagher_widespread_2018}, that star formation might be triggered inside galactic outflows. Such result would imply the existence of a different mode of star formation, which may play an important role in our understanding of the evolution of galaxies, for instance contributing to the assembly of stellar mass at high redshift. 

The identification of regions in galaxies affected by feedback processes can be done in different ways, including the detection of outflowing gas moving at relatively high velocities with respect to the systemic components \citep[e.g.,][]{westmoquette_studying_2008}, the study of excitation maps of shock-ionized gas \citep[e.g.,][]{veilleux_identification_2002}, or by exploring their effect on the magnetic fields of galaxies \citep{damas-segovia_chang-es._2016}. These outflows have been observed in all gas phases: neutral \citep{cazzoli_spatially_2014, rubin_evidence_2014}, molecular \citep{cicone_massive_2014, pereira-santaella_spatially_2018} and ionized \citep{westmoquette_studying_2008, arribas_ionized_2014}. The study of galactic outflows has greatly benefited from the advent of Integral Field Spectroscopic (IFS) instruments, given that they allow spatially-resolved studies of the kinematics, ionization mechanisms, morphologies and physical properties of the outflowing gas across entire galaxies \citep{bellocchi_vlt/vimos_2013, cazzoli_spatially_2014}. In particular, large IFS surveys such as CALIFA \citep{sanchez_califa_2012}, SAMI \citep{allen_sami_2015} and MaNGA \citep{bundy_overview_2015} offer the possibility of exploring the incidence and properties of outflows in a large number of galaxies, allowing a statistically meaningful characterization of such phenomenon. 

In this work we study the incidence and properties of ionized outflows in a large number of galaxies ($>$ 2700) observed within the MaNGA survey \citep[Mapping Nearby Galaxies at Arecibo Point Observatory;][]{bundy_overview_2015} and included in their Data Release 2 \citep[DR2;][]{abolfathi_fourteenth_2018}. This data set includes galaxies with a wide range of stellar masses, star formation rates ($SFRs$), AGN activity, etc., that allows the chracterization of the kinematics and physical properties of the ionized gas in regions hosting ionized outflows. We also investigate the role that outflows might play regulating the star formation activity and metal content of galaxies.
A related study on ionized outflows based on the same sample has been recently published by \citet{gallagher_widespread_2018}, where they focus on the study of star formation taking place within ionized outflows and its implications in galaxy evolution. 

In Section~\ref{sec:data} we describe the MaNGA DR2 data; Section~\ref{sec:analysis} contains the different steps of our analysis, from the MaNGA datacubes to the spectral modelling of the emission-line spectra; in Section~\ref{section:sample_selection} we explain the criteria applied to identify ionized outflows;  Section~\ref{sec:results} contains the main results from our analysis of the kinematics and properties of the ionized gas, including also discussions about the implications of our results in the context of galaxy evolution; finally, in Section~\ref{sec:summary} we present the summary and main conclusions of this work. Throughout this work we adopt a cosmology with $H_{\rm0}$~=~67.3~kms$^{-1}$~Mpc$^{-1}$, $\Omega_{\rm M}$~=~0.315, $\Omega_{\Lambda}$~=~0.685 \citep{planck_collaboration_planck_2014}.

\section{Data sample}
\label{sec:data}
This work is based on data from the second Data Release of the MaNGA survey \citep[DR2;][]{abolfathi_fourteenth_2018}, one of the three core programs of Sloan Digital Sky Survey IV (SDSS-IV). All the galaxies in the MaNGA sample are part of the main SDSS parent sample and are selected to have stellar masses, $M_{\star}$, above $10^{9}$ M$_{\odot}$ and redshifts, $z$, up to 0.15. The spectra span approximately from $3600$~\AA{} to $10000$~\AA, with a spectral resolution that ranges from R$\sim1400$ at $4000$~{}\AA~to R $\sim2600$ at $9000$~\AA, corresponding to $\sim90$~km/s and $\sim49$~km/s, respectively. The Point Spread Function (PSF) in the datacubes is 2.5 arcsec (FWHM). 

The MaNGA DR2 parent sample is divided into different subsamples. For our analysis we select all the galaxies that are part of the so-called Primary, Secondary and colour-Enhanced samples. The two former ones are designed to reach 1.5~$R_{\rm e}$ and 2.5~$R_{\rm e}$ (being $R_{\rm e}$ the effective radius) in $80\%$ of their targets, respectively, whereas the last one is designed to include under-represented regions of the $M_{\star}$-colour space (see \citet{wake_sdss-iv_2017} for more details on the different samples). The combination of these three samples makes up a total of 2672 galaxies. In this work we use stellar mass estimates, when available, from the SDSS DR7 \citep{kauffmann_stellar_2003}. 


\section{Analysis}
\label{sec:analysis}

\subsection{From datacubes to \Ha maps}
\label{processingcubes}
\begin{figure}
\begin{center}
\includegraphics[width=0.48\textwidth]{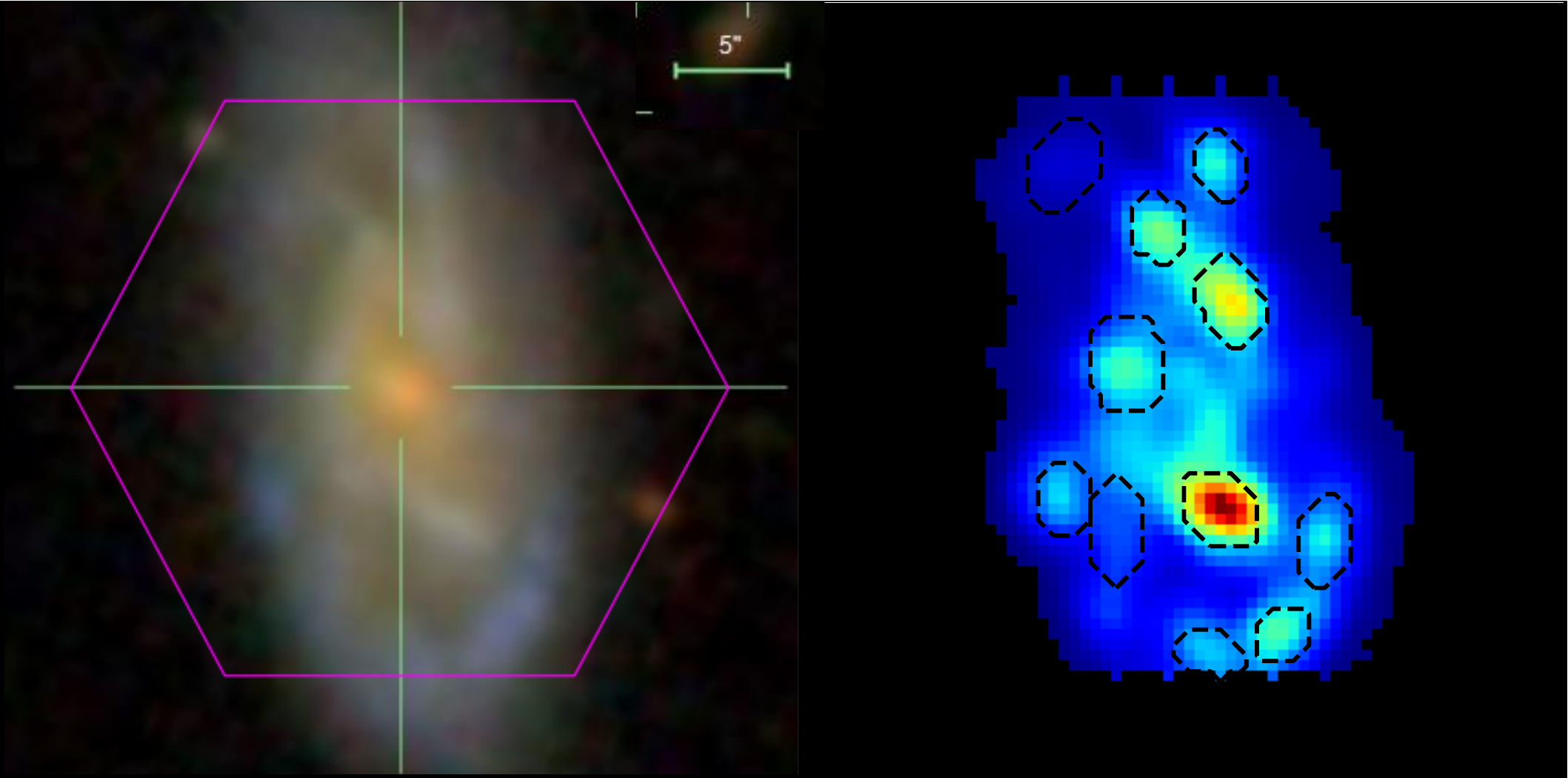}
\caption[example Halpha clumps]{Example of the \Ha clumps identification in a galaxy. \emph{Left}: Field of view of MaNGA (pink hexagon) superimposed on the SDSS coloured image of the galaxy. \emph{Right}: Map of the \Ha flux distribution showing the regions identified as \Ha-emitting clumps.}
\label{Halphamap}
\end{center}
\end{figure}

In this work we use the MaNGA datacubes with linear wavelength sampling (LINCUBE). From the datacubes we use the three-dimensional flux and inverse variance per unit wavelength, a bad pixel mask and the spectral resolution as a function of wavelength. Using these data we generate 
H$\alpha$ maps for all the galaxies including only the spaxels with $S/N > 3$ in the \Ha line, which is estimated as in \citet{rosales-ortega_integrated_2012}. In this procedure we discard all the spaxels where the spectra around the target emission lines are affected by bad pixels. In this way, we generate maps for $\sim1500$ galaxies showing clear \Ha emission, roughly 55\% of the original MaNGA DR2 sample. An example of one of the \Ha flux maps generated is shown in Figure \ref{Halphamap}.


\subsection{Identification of the \Ha-emitting regions}
\label{sec:regions}

Once the \Ha maps are generated and, given the clumpy distribution of the \Ha emission, we develop a method to identify individual regions characterized by \Ha emission across the galaxies. We note that the corresponding size of the MaNGA PSF in kpc ($\sim1.5$~kpc at the median redshift of the sample, varying from $\sim0.6$~kpc to $\sim6.73$~kpc) prevents us from identifying individual star-forming regions, which typically have sizes of $20-100$~pc \citep[e.g.,][]{miralles-caballero_extranuclear_2012}. Therefore, the regions identified here probably contain several star-forming regions in most of the cases.  

In order to isolate the \Ha-emitting regions across the MaNGA coverage of the galaxies in an automated way, we start by finding the peaks of emission at different intensity levels using the algorithm CLUMPFIND \citep{williams_determining_1994}. This algorithm searches for local peaks of emission and follows them down to lower intensity levels, decomposing the \Ha maps into a set of independent structural units or `clumps' where the emission is concentrated. Once we have identified the \Ha-emitting regions, we define the sizes of the apertures to extract the spectra by fitting a two-dimensional elliptical isophote with a Gaussian profile. In this procedure both minor and major axes of the elliptical isophote have to be larger than the FWHM of the MaNGA PSF (2.5 arcsec). For each of the regions, we produce an integrated spectrum by combining all the spaxels with a $S/N > 3$ in the \Ha line within the chosen aperture. We do not align the spectra from these spaxels to correct for the systemic velocity of the galaxy in order to preserve the shape of a possible broad component. Finally, given that the MaNGA datacubes do not include covariance calculations, when combining the spectra we calibrate the noise vectors following the recipes from \citet{law_data_2016}. As a result of this analysis, we identify a total of $\sim5200$ \Ha-emitting regions across the galaxies, both nuclear and off-nuclear, with a mean of $\sim3.5$ regions per galaxy.




\subsection{Subtraction of stellar component}
\begin{figure}
	\begin{center}
		\includegraphics[width=0.49\textwidth]{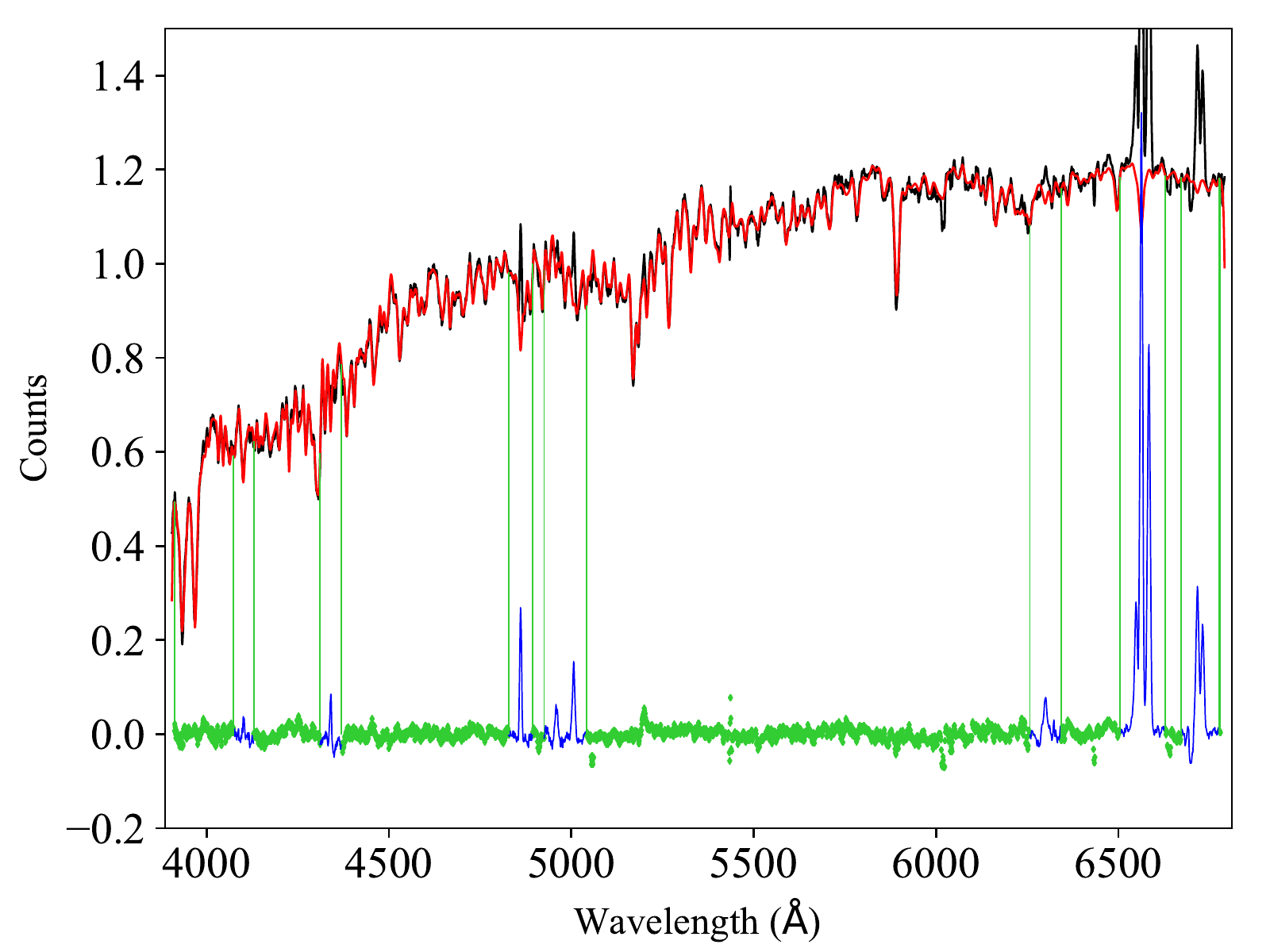}
		\caption[pPXF example]{Example of the stellar continuum fit to the spectrum of one of the \Ha-emitting regions using the software pPXF \citep{cappellari_improving_2017}. The \emph{black} line corresponds to the observed spectrum whereas the \emph{red} line is the best fit to the data. The residuals are shown as a \emph{green} line and the emission lines, that are masked during the fitting procedure, are shown in \emph{blue}.}
		\label{ppxf_example}
	\end{center}
\end{figure}

In order to constrain the kinematics and properties of the ionized gas in the \Ha-emitting regions we need to correct their  emission-line spectra for the absorption produced by the stellar populations inhabiting them. To do that, we use the software pPXF \citep[Penalized Pixel-Fitting;][]{cappellari_improving_2017}, a tool that models the stellar continuum in the spectra using a maximum penalized likelihood approach. To fit the stellar contribution we use the PEGASE-HR simple stellar population (SSP) models \citep{le_borgne_evolutionary_2004}, which have the spectral resolution (FWHM$\sim$0.5~\AA) and wavelength coverage ($3900 - 6800$~\AA) appropriate to fit the MaNGA spectra. In this procedure we take into account the wavelength variation of the spectral resolution in the MaNGA spectra. After fitting the stellar continuum, we subtract it from the galaxies' spectra to produce pure emission-line spectra. An example of the performance of pPXF is shown in Figure \ref{ppxf_example}. 

\subsection{Spectral fitting}
\label{section:spectral_fitting}

The spectra from regions hosting ionized outflows are characterized by the presence of a secondary, broad kinematic component associated to high gas velocities (see Section \ref{section:sample_selection}), superimposed on the systemic emission from the host. When present, this signature is more easily detected in strong emission lines such as \Ha or \OIIIb. In these cases, the properties of the outflowing gas can be constrained by fitting the emission-line spectra with a model composed of two Gaussian profiles, corresponding to the narrow (systemic) and broad (outflow) kinematic components. For strong outflows, such as those observed in AGN or starburst galaxies, the broad component can be easily detected and characterized. However, this task becomes significantly more difficult when the outflow is intrinsically weak compared to the total emission from the host region or when the geometry of the outflow or the host galaxy (such as close to edge-on inclinations) hinder the detection of the broad emission. Moreover, since the emission lines associated to the outflow are intrinsically broad, the flux is spread across more spectral pixels and the $S/N$ can be low. 

We perform the spectral fitting using a Bayesian approach based on Markov Chain Monte Carlo (MCMC) techniques. In particular, we employ the software developed by \citet{foreman-mackey_emcee_2013}, \textsc{emcee}, a tool that implements the Affine Invariant MCMC Ensemble sampler of \citet{goodman_ensemble_2010}. Using this method, we explore the whole parameter space by sampling more intensively the regions of high likelihood but at the same time allowing the exploration of regions of lower likelihood, preventing in this way the method from getting trapped in regions of local maxima. 

For the purpose of our study we simultaneously fit a model consistent of two sets of Gaussian functions (associated to the narrow and broad kinematic components) to the emission lines  \Ha, \Hb, \NIIb, \OIIIb, \SII$\lambda\lambda6717,6731$. In the fitting procedure, we apply the same kinematics (velocity and velocity dispersion) to all the emission lines, fixing the ratio between the two \NII$\lambda\lambda6548,6583$ lines to $3.06$ \citep{osterbrock_astrophysics_1989}. In total, the parameters included in the model are the redshift ($z$) and width ($\sigma$) of each set of Gaussians, the fluxes of each kinematic component in the six spectral lines and the continuum around them (intercept and slope). To initialize the MCMC sampler we provide the following priors (i.e., the constraints we set on the parameter values based on our previous knowledge about them): the fluxes and widths of the lines have to be positive (our spectra have been selected to show clear emission lines), the value of the \Ha/\Hb ratio larger or equal than 2.86 (case B recombination value), and the ratio between \SII$\lambda\lambda6717,6731$ in the range [$0.44-1.42$] \citep{osterbrock_astrophysics_1989}. We convolve each emission line with the corresponding instrumental profile at a given wavelength, which is provided in the header of the MaNGA datacubes. 

\begin{figure}
\begin{center}
\includegraphics[width=0.49\textwidth]{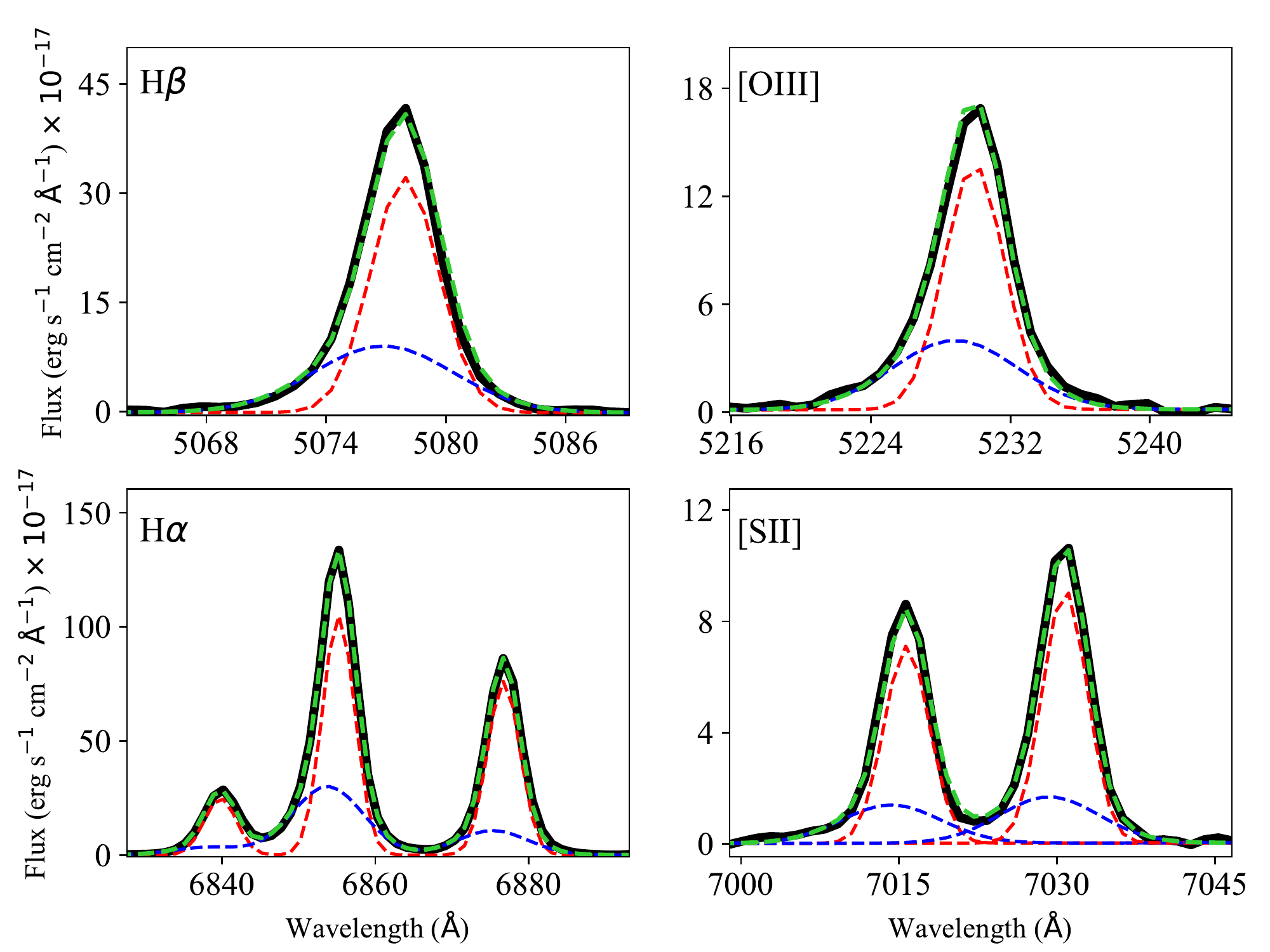}
\caption[2comp fit]{Example of the results from the MCMC spectral modelling for one of the \Ha-emitting regions in the MaNGA datacubes showing a blue-shifted kinematically distinct component, likely associated to an outflow. The \emph{solid, black} line is the observed spectrum, the \emph{dashed-red} and \emph{dashed-blue} lines correspond to the narrow and broad kinematic components, respectively, and the \emph{dashed-green} line is the sum of these two components. The values of the parameters for the model represented here correspond to the median of their posterior probability density distributions (see Section \ref{section:spectral_fitting}).}
\label{plot_2compfit}
\end{center}
\end{figure}

As output from the MCMC fitting we obtain the posterior probability distributions for all the model parameters. Throughout this paper we use the median values from these distributions whereas the errors on these values are estimated using the 16th and 84th percentiles. In Figure~\ref{plot_2compfit} we show an example of the spectral modelling to the spectrum of one of the \Ha-emitting regions studied here, in this case showing asymmetries in the line profiles indicating that a blue-shifted, ionized outflow is probably present in that region. We repeat the previous spectral fitting using a model with a single kinematic component. The result from this additional fit is used in the next section to discern when an outflowing component is present in the spectra.


\begin{figure*}
	\begin{center}
		\includegraphics[width=0.99\textwidth]{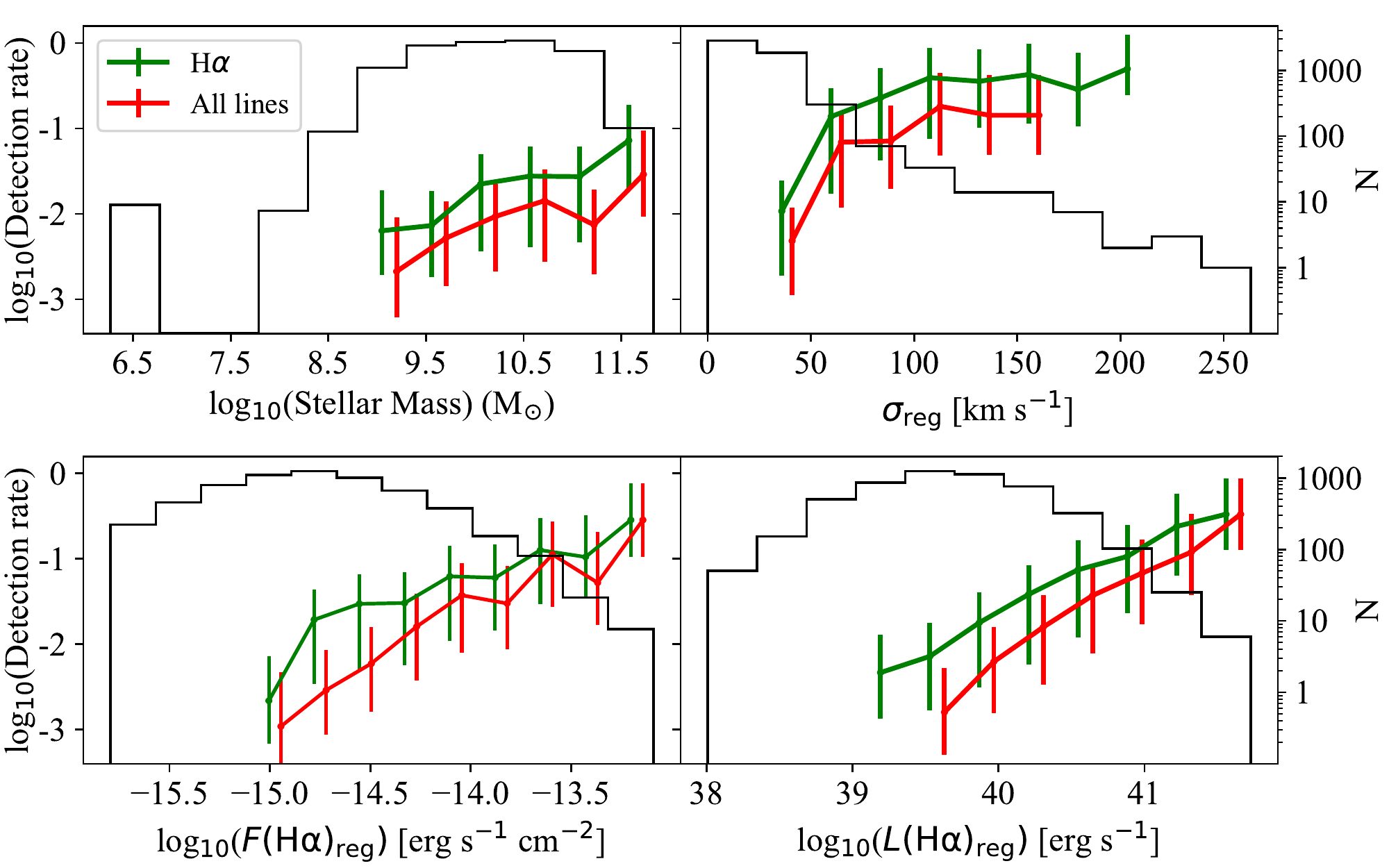}
		
		\caption[]{Detection rate of outflows as a function of the stellar mass of the host galaxy and properties of the individual host regions: velocity dispersion of the gas, total \Ha flux and total \Ha luminosity. We represent the detection rate for the sample of outflows detected in \Ha $\emph{green}$ and the sample of outflows detected in all the emission lines $\emph{red}$. The $\emph{black}$ histograms correspond to the parent sample containing all the individual galaxies (upper-left panel) and \Ha-emitting regions (other three panels).}
		\label{plot:histo_lumha_masses}
	\end{center}
\end{figure*}

\section{Sample selection of outflow candidates}
\label{section:sample_selection}

Similarly to previous works \citep[e.g.,][]{shapiro_sins_2009, genzel_sins_2011, soto_emission-line_2012, arribas_ionized_2014, gallagher_widespread_2018}, we consider evidence of outflowing gas only when an extra broad component is clearly identified and it implies  velocities \emph{significantly larger} than those corresponding to the systemic motions in the selected SF regions, as traced by the narrow component.  In more detail, we consider the presence of ionized outflows when the following criteria are fulfilled:

\begin{itemize}
\item {\bf Detection of two clear components in emission:} we select the cases where both kinematic components have a probability greater than 99.7\% of being detected with $S/N>~3$ integrated in the line, and the contribution of the weakest component to the total line flux is $>5\%$. 
\\
\item {\bf Presence of radial outflowing gas:} we consider that an outflowing component is present when $\sigma_{\rm broad}>$~$~1.4~\times$~$\sigma_{\rm narrow}$\footnote{In a few cases where the lines are barely resolved (i.e., the line width is close to that of the Line Spread Function), we ensure that $\sigma_{\rm broad}$ is also $1.4\times$ larger than the instrumental broadening, $\sigma_{\rm inst}$, at the given wavelength. In these cases, we visually inspect the spectra to check that the classification is adequate.} and the broad kinematic component is substantially broader than the one obtained in the single kinematic component fit ($\sigma_{\rm broad}>$~$~1.2~\times$~$\sigma_{\rm 1comp}$).
\end{itemize}


Finally, we discard a few cases where the spectra require additional kinematic components, indicative of the presence of type 1 AGN. 

We note that as a consequence of applying these criteria the outflows we detect have a minimum $FWHM_{\rm broad}$ of $\sim200$~km/s.


This method for identifying outflows is similar to other ones used in the literature \citep[e.g.,][]{ho_sami_2014, schreiber_kmos^3d_2018, gallagher_widespread_2018}. In fact, in the work by \citet{gallagher_widespread_2018}, where they also used MaNGA data, they found that the broad component is associated to outflowing gas and is characterized by similar FWHMs to the ones we find in our work ($>150$km/s; see Figure~\ref{plot:FWHM_vs_LHa} in this paper and Figure~3 in \citet{gallagher_widespread_2018}).

Using these criteria we define two samples: one for which we require the two components to be detected at least in the \Ha line, which contains 105 \Ha-emitting regions, and a more robust one were the two components have to be detected in all the emission lines included in the spectral fitting, i.e., \Ha, \Hb, \NIIb, \OIIIb, \SII$\lambda6717,6731$, this latter one containing 45 \Ha-emitting regions. The motivation for using a sample based only on the \Ha emission is the study of the kinematic properties of the outflow component down to when the contribution is intrinsically weak and/or dust attenuation or geometry (i.e. inclination) hampers the detectability in other lines. We will use the second sample to better characterize the properties of the ionized gas in the outflowing component. 

During the analysis of the MaNGA datacubes, we serendipitously detected emission from the supernova candidate 2015co in a galaxy at $z\sim0.029$, whose supernova nature was confirmed and reported in \citet{rodriguez_del_pino_transient_2018}. 






\section{Results and Discussion}

\label{sec:results}

\subsection{Outflows detection rate} 
\label{sec:incidence}

We start by identifying the outflows that are located in regions away from the central parts of their hosts. Such task is difficult in many cases due to the clumpy nature of the \Ha emission, limited by the spatial resolution of the instrument, particularly in galaxies with high inclination and at high redshift. Therefore, we only consider as off-nuclear outflows the cases when the broad emission originates in a region that can be spatially isolated from the emission coming from the nuclear parts of the galaxies. Following this method, out of the 105 ionized outflows detected in the \Ha emission, we identify 2 of them in off-nuclear regions at projected distances larger than 1~kpc from the galaxies' centers. Unfortunately, none of them is detected in other emission lines apart from \Ha (see Figure~\ref{plot:plot_8highest_fwhm}). Such low fraction of off-nuclear outflows indicates that most of the outflows tend to originate in the nuclear parts of galaxies. Within the selected sample, we find that a single galaxy is host of three regions showing outflow components (MaNGA~8588-6101), located along a star-forming ring around a central AGN. Although the three regions are located in different regions of the galaxy, the effects from the AGN cannot be discarded; thus, we will not consider them as off-nuclear outflows.

Overall, we detect ionized outflows (at least in \Ha) in $\sim2$\% of the \Ha-emitting regions and in $\sim7\%$ of the \Ha-emitting galaxies studied in this work. 

We explore the outflows detection rate as the fraction of \Ha-emitting regions showing outflows for a given local value of the systemic velocity dispersion, $\sigma_{\rm reg}$, the local \Ha flux and luminosity (including the contribution from the two components when detected) and a given global value of stellar mass. For this analysis, we show in Figure~\ref{plot:histo_lumha_masses} the outflows detection rate as a function of the different parameters for the two samples of regions hosting outflows that we have defined in Section~\ref{section:sample_selection}. We also include the histograms with the distribution of the different parameters for the parent sample of \Ha-emitting regions.

\begin{figure}
	\begin{center}
		\includegraphics[width=0.45\textwidth]{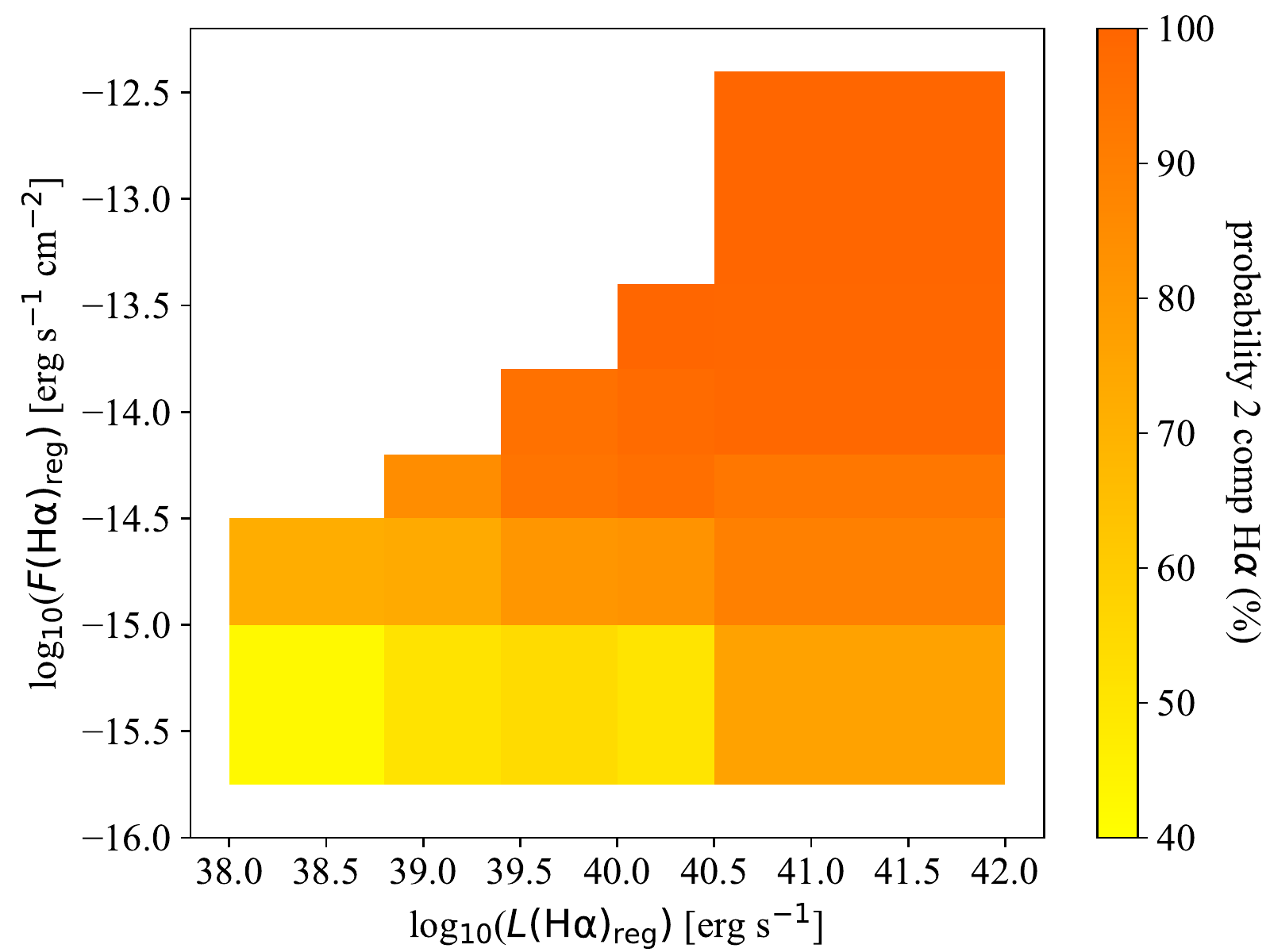}
		\caption[selection_effects]{Sample of all the individual H$\alpha$-emitting regions studied in this work distributed in different bins of total \Ha flux and luminosity (including the contribution from the two components when detected). The bins are color-coded according to the average probability of detecting two emission line components in H$\alpha$, with $S/N > 3$. The bins are chosen to include at least 10 H$\alpha$-emitting regions, therefore covering different parameter ranges (shapes).}
		\label{plot:selection_effects}
	\end{center}
\end{figure}


We find a clear increase in the detection rate of outflows towards higher \Ha fluxes and luminosities (Spearman's correlation test probabilities $>3\sigma$). In particular, with respect to flux their detection rate increases from $2.6\%_{-0.3}^{+0.3}$ at fluxes below $\sim1.0\times10^{-14}$~erg~s$^{-1}$cm$^{-2}$ to $9.3\%_{-1.5}^{+2.2}$ in regions with fluxes above that value; with respect to luminosity, their detection rate increases from $1.0\%_{-0.2}^{+0.2}$ at L(\Ha)$\leq10^{40}$~erg~s$^{-1}$ to $5.5\%_{-0.6}^{+0.7}$ for higher luminosities. The increase towards higher gas velocity dispersions and stellar masses are less pronounced but still significant (Spearman's correlation test probabilities $>2\sigma$). With respect to gas velocity dispersion there seems to be a flattening in the detection rate of outflows above $\sigma_{\rm reg} >$100kms$^{-1}$, whereas regarding mass, their detection rate increases from $1.0\%_{-0.2}^{+0.3}$ at $\leq10^{10}$M$_{\odot}$, to $\sim2.9\%_{-0.3}^{+0.3}$ for higher masses.
We detect outflows in regions with \Ha fluxes down to $\sim1.0\times10^{-15}$~erg~s$^{-1}$cm$^{-2}$, \Ha luminosities as low as $\sim1.1\times10^{39}$~erg~s$^{-1}$ and in host galaxies with stellar masses above $\sim1.2\times10^{9}$M$_{\odot}$. 

It is important to note here that we detect ionized outflows at low \Ha luminosities, corresponding to $SFRs$ down to $\sim0.01$~M$_{\odot}$yr$^{-1}$ (see Figure~\ref{plot:plot_3x3_kinematics}), a regime that has only been explored using stacking of spectra from thousands of galaxies \citep{cicone_outflows_2016}. Therefore, with our analysis we are able to probe ionized outflows down to low $SFRs$ for individual objects, extending the ranges explored by previous works \citep[e.g.,][]{rupke_outflows_2005, arribas_ionized_2014, heckman_implications_2016}.



\begin{figure*}
	\begin{center}
		\includegraphics[width=0.98\textwidth]{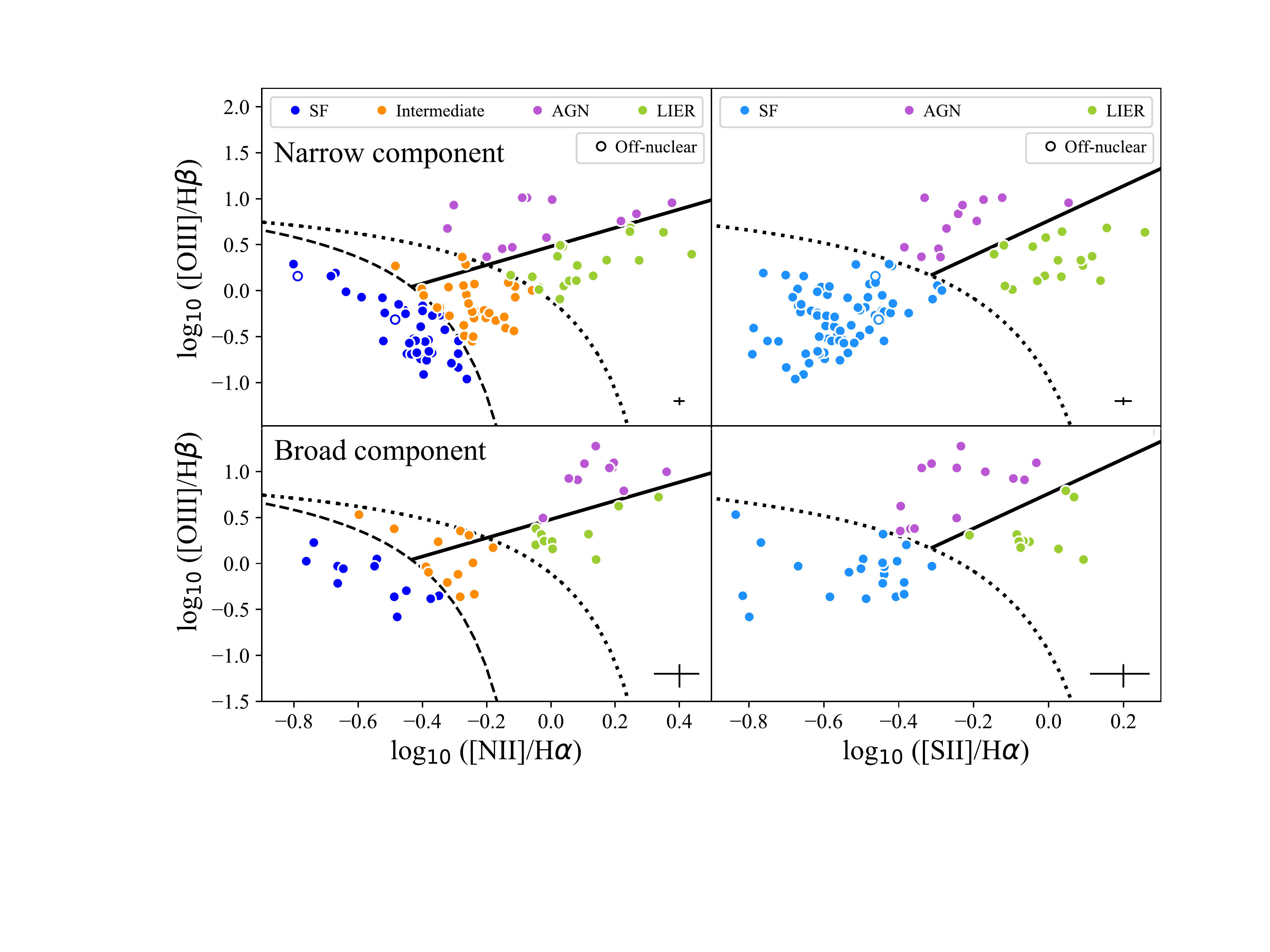}
		\caption[bpt_clumps]{BPT diagnostic diagrams for the narrow (\emph{top}) and broad components (\emph{bottom}) in the regions where ionized outflows are detected at least in the \Ha line. The upper panels contain all 105 the regions hosting outflows (detected in the \Ha line) whereas the bottom panels contain only the 45 regions where outflows are detected in all the emission lines (see Section~\ref{section:sample_selection}). The line fluxes used for the upper plots correspond to the systemic (narrow) component or, in the case that the outflow is not detected in all the lines, to the values obtained in the single component fit. The different lines shown in both figures are the empirical separations between ionizing sources of different types, as described in the legend (see Section \ref{section:bpt} for more details). Typical (mean) error values are shown in the bottom-left corner of the panels.}
		\label{plot:bptplots}
	\end{center}
\end{figure*}

	Although the detection rate of ionized outflows is low, we have to bear in mind that non detections do not imply their absence. The detectability of outflows might depend on different factors such as the extinction associated to the outflowing component (as we explore later in this work), the amount of gas in the surroundings and the efficiency at which the energy is injected into it. Another factor that has been found to influence the detectability of outflows, at least in the neutral phase of the gas, is the inclination of the galaxy \citep{heckman_absorption-line_2000}. To test this dependency in our sample we estimate the inclinations of the MaNGA DR2 galaxies in the same way as described in \citet{chen_absorption-line_2010}: using the axis ratio, $b/a$, and the $r$-band absolute magnitude, $M_{\rm r}$, of the galaxies (which are taken from the MPA/JHU Value Added Catalogue\footnote{\url{http://www.mpa-garching.mpg.de/SDSS/DR7}}) in combination with Table 8 from \citet{padilla_shapes_2008}. In the same way as they do, we only consider the inclinations for disk galaxies, which are selected by requiring the parameter $fracDeV$ to be $<$ 0.8 ($fracDeV$ quantifies the profile type based on the SDSS best linear combination of an exponential and de Vaucouleurs models). Comparing the inclinations of galaxies in the sample with detected outflows with those in the parent sample, a Kolmogorov-Smirnov test gives a $p$-$value\sim0.21$, indicating that we cannot reject the hypothesis that the two distributions are the same. The median inclination of galaxies with detected outflows is $i~\sim50\degree$, similar to that of the parent sample of galaxies with \Ha emission, $i \sim55\degree$. These two values are also close to the median inclination of the initial sample of 2672 MaNGA DR2 galaxies considered in this study (see Section \ref{sec:data}), which is also $i \sim55\degree$. Thus, we do not find that galaxies hosting ionized outflows have higher inclinations than the rest of the galaxies in our sample. However, we have to bear in mind that due to the generally low incidence of outflows detected any trend with inclinations could be blurred out. On top of that, although the MaNGA sample was not selected based on inclination \citep{wake_sdss-iv_2017}, the relatively high inclination of the MaNGA DR2 galaxies ($i \sim55\degree$) could also hamper the study of inclination effects. 

Finally, we explore whether the decline in the outflows detection fraction at low \Ha fluxes is due only to selection effects (i.e., higher difficulty in detecting two components in fainter lines). We note that trying to establish whether the lack of detections is due to a low observed flux or to a low intrinsic luminosity is difficult because these two quantities are intrinsically correlated. 
To evaluate the presence of selection effects we study the probability (obtained from the MCMC modelling) of detecting two emission line components as a function of \Ha flux and luminosity. This is illustrated in Figure~\ref{plot:selection_effects}, where we have distributed all the individual H$\alpha$-emitting regions in different bins of \Ha flux and luminosity (including the contribution from the two components when detected) and estimated the average probability of detecting two emission line components in H$\alpha$ with $S/N > 3$. As seen in the plot, the probability increases as a function of the observed flux, indicating selection effects, but also as a function of the intrinsic luminosity, implying that the detectability of outflows is also linked to the intrinsic properties of the host regions. Therefore, the observed decrease in the detection rate of outflows for lower \Ha fluxes is not only due to selection effects but also to intrinsic lower luminosities.





\subsection{Identifying the ionizing source in the host regions}
\label{section:bpt}
There is a wide variety of physical mechanisms that can give rise to ionized outflows, such as AGN activity, supernova explosions, stellar winds and shocks. The differences between the various mechanisms can be primarily quantified using two factors: the kinetic energy injected into the gas and the level of ioinization of the gas. The former one can be traced by the kinematics associated to the outflowing component, whereas the latter one can be constrained by mean of standard BPT diagnostic diagrams, \citep{baldwin_classification_1981}. 
In this section we are interested in characterizing the ionization in the regions hosting outflows, i.e., the systemic (or narrow) kinematic component; therefore, whenever the broad kinematic component is only detected in \Ha, for the other lines we
adopt as systemic value the fluxes measured in the single component fit. 

In the top panels of Figure~\ref{plot:bptplots} we show two versions of the BPT diagram for our sample of regions hosting ionized outflows. On the one hand, the BPT-N diagram (left panel) separates the sources in four different populations by using a set of different empirical lines: the dashed line that is used to isolate star-forming regions is the one proposed by \citet{kewley_theoretical_2001}, whereas the dotted line that separates AGN and LIERs is taken from \citet{kauffmann_host_2003}. The area in between these two lines is populated by the so-called `intermediate' population. Finally, the solid line is aimed at separating AGN from LIERs, as suggested by \citet{cid_fernandes_alternative_2010}. The `intermediate' population corresponds to regions where the level of ionization cannot be explained by either star formation or AGN activity alone \citep{cid_fernandes_alternative_2010}, therefor its location in the BPT diagram. First introduced as Low-Ionization Nuclear Emission Regions (LINERs) by \citet{heckman_optical_1980} because they were associated to nuclear emission, LIERs \citep{monreal-ibero_liner-like_2006, belfiore_sdss_2016} are regions, that can be extended and located away from the nucleus, hosting low levels of ionization that can be produced by low-luminosity AGN, fast shocks, starburst-driven winds or diffuse ionized plasma \citep{collins_ionization_2001, dopita_spectral_1995, armus_optical_1990}. On the other hand, the BPT-S diagram (right panel) separates the sources in three types SF, AGN and LIERs using the lines suggested by \citet{kewley_theoretical_2001} and \citet{kewley_host_2006}. 

\begin{figure*}
	\begin{center}
		\includegraphics[width=0.98\textwidth]{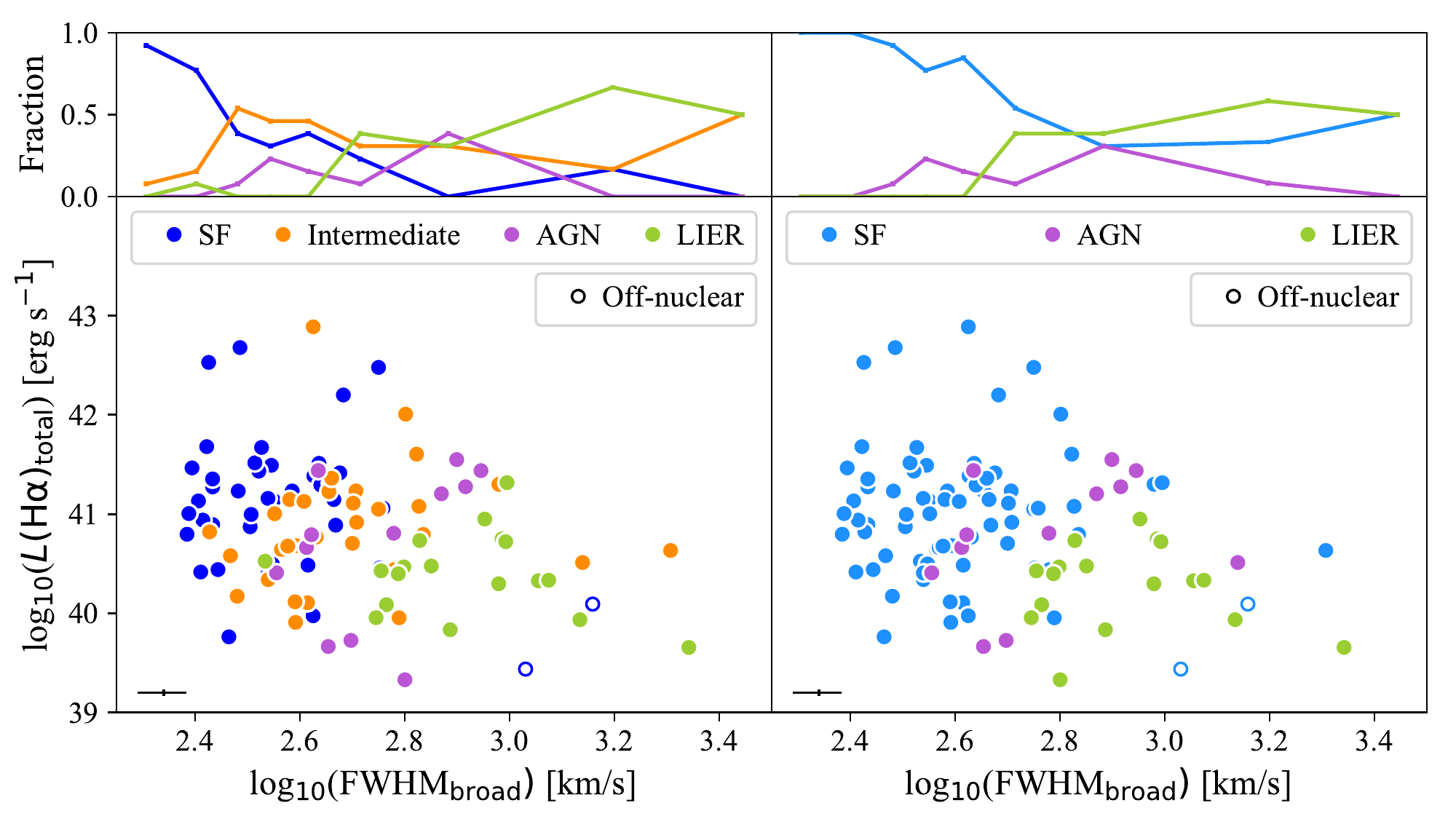}
		\caption[FWHM vs Halpha luminosity]{\emph{Bottom panels}: comparison of the total \Ha luminosity and the kinematics associated to the ionized outflows detected in the \Ha, as traced by $FWHM_{\rm broad}$. The colours indicate the type of source that is ionizing the gas in the host region, as found in the two BPT diagnostic diagrams shown in Figure \ref{plot:bptplots}. Typical (mean) error values are shown in the bottom corner of the two panels. \emph{Top panels}: Distributions of $FWHM_{\rm broad}$ values for the different types of sources. }
		\label{plot:FWHM_vs_LHa}
	\end{center}
\end{figure*}

\begin{table}
\begin{center}
\label{table:BPT_incidence}
\begin{tabular}{lcc}
\hline
\hline
 & BPT NII & BPT SII\\
\hline

AGN & 41.2 \% & 18.2 \% \\ 
LIER & 39.0 \% & 33.8 \%\\   
Star-forming & 1.0 \% & 1.4 \%\\
Intermediate & 9.4  \% & - \\

\hline

\end{tabular}
\caption{Detection rates of outflows as a function of the type of ionizing source in the host regions, classified using the two BPT diagnostic diagrams presented in Section~\ref{section:bpt} and Figure~\ref{plot:bptplots}.}

\end{center}
\end{table} 

Following these classifications, we find ionized outflows associated to all possible types of ionization, indicating the variety of conditions in which outflows can originate. Based on the results from the two BPT diagnostic diagrams, ionization from star formation is found in 39\% of the galaxies using the BPT-N diagram and in 71\% of the galaxies using the BPT-S one. The fractions of sources with ionization coming from AGN and LIERs are $\sim11\%$ and $\sim17\%$, respectively, being similar in both diagrams. From all the sources classified as `intermediate' in the BPT-N diagram, only one is not classified as star-forming in the BPT-S, corresponding to an AGN located close to the dividing line of the SF population. The two off-nuclear regions hosting outflows identified in this work are classified as star-forming in both diagrams.

{To complement the analysis presented in the previous section, we evaluate now the detection rate of outflows as a function of the BPT class. We have classified all the regions in the parent sample where all the relevant emission lines are detected with a $SN>5$ ($>98\%$ of the regions) using both BPT diagnostic diagrams. For this analysis, when two components are clearly detected (Section~\ref{section:sample_selection}) we adopt the $S/N$ in the narrow kinematic component. The detection rates are presented in Table~\ref{table:BPT_incidence}. Our results show that the highest detection rates are associated to regions characterized by AGN and LIER emission, whereas for star-forming regions, which are the largest number in the sample, the detection rates are much lower.

Although both BPT diagrams provide relevant information about the ionization mechanisms in our sources. Given that both diagrams select  almost the same sources as AGN (only one differing object) and LIERs (two differing objects), throughout the paper we will use the BPT-N diagram because it allows a separate study of the SF and `intermediate' populations. 


\subsection{Outflow properties}

\subsubsection{Kinematics of the ionized gas.}
\label{section:kinematics}

Once we have classified all the regions hosting ionized outflows, we explore the kinematic differences between the outflows originated in them. In Figure~\ref{plot:FWHM_vs_LHa} we show the FWHM of the broad kinematic component, $FWHM_{\rm broad}$, as a function of the total \Ha luminosity, obtained as the sum of the \Ha fluxes in the narrow and broad kinematic components, for the different populations identified in the BPT diagrams. The large scatter indicates that the outflowing gas can have a wide range of velocity dispersions (i.e., FWHM), ranging from $\sim200$ km/s up to $\sim2500$km/s in the extreme cases. Apart from this scatter, there are differences in the $FWHM_{\rm broad}$ for outflows associated to different types of ionization. On average, the regions with SF ionization have outflows with the lowest $FWHM_{\rm broad}$, with a median of $\sim350$ (390) km/s in the BPT-N (BPT-S) classification, although these regions can also host outflows with extreme kinematics ($\geq1000$~km/s). The median $FWHM_{\rm broad}$ associated to outflows increases progressively for regions classified as `intermediate', ($\sim450$~km/s), and those classified as AGN ($\sim550$ km/s) and LIERs ($\sim900$ km/s), which show similar distributions in both BPT diagrams. These results indicate that there is a clear distinction between the kinematics of outflows originated in regions ionized by different mechanisms, although the ranges of \Ha luminosities spanned are very similar. We find no significant correlations (Spearman's correlation tests yield probabilities less than $2\sigma$) between \Ha luminosity and $FWHM_{\rm broad}$ for any of the populations identified in the figures. As examples, in Figure~\ref{plot:plot_8highest_fwhm} we show the spectral fit to the ionized outflows with $FWHM_{\rm broad}$~$>$~$1000$~km/s.  

	In their study of neutral outflows in a sample of 78 starburst galaxies, \citet{rupke_outflows_2005} also found that outflows in LIERs tend to have higher FWHMs than those associated to HII galaxies, 373~km/s vs~253 km/s (median values after converting from their Doppler parameter $b$). However, the $FWHM_{\rm broad}$ we measure in the ionized outflows are significantly larger than the ones estimated by \citet{rupke_outflows_2005} for the neutral gas phase, $\sim110\%$ and $\sim50\%$ for LIERs and SF galaxies, respectively, a difference that they reported to be $\sim70\%$.
	
Interestingly, the outflows in the two off-nuclear, star-forming regions, have $FWHM_{\rm broad}$~$\geq$~1000~km/s, with \Ha luminosities $1.86\times10^{39}$~erg s$^{-1}$ and $2.74\times10^{40}$ erg s$^{-1}$. Such high kinematics in off-nuclear regions are very unusual, with only a few similar cases been reported in the literature \citep[e.g.,][]{castaneda_remarkable_1990, gonzalez-delgado_violent_1994}. These events are thought to be associated with a superbubble in the blow-out phase, produced by a large number of massive stars and/or supernovae \citep{roy_origin_1992}. A detailed study exploring the characteristics of these two sources and their possible origin will be presented in a separate paper (Rodriguez Del Pino et al. in prep.).

We extend the study of the kinematic properties of the outflowing gas in Figure~\ref{plot:plot_3x3_kinematics} where we show the values of $FWHM_{\rm broad}$, $\Delta V$ and $V_{\rm max}$ as a function of $L$(H$\alpha$), $\Sigma_{L({\rm H}\alpha)}$ and stellar mass of the host galaxy, when available in SDSS DR7, \citep{kauffmann_stellar_2003}. The values of $\Sigma_{L({\rm H}\alpha)}$ are estimated using the areas of the \Ha-emitting regions defined in Section~\ref{sec:regions}. Given that the sizes we estimate ($> 1-6$~kpc) are larger than those typical of star-forming regions \citep[$20-100$~pc;][]{miralles-caballero_extranuclear_2012}, the values of $\Sigma_{L({\rm H}\alpha)}$ estimated here should be considered lower limits. The parameter $V_{\rm max} =  |\Delta V| + FWHM_{\rm broad}/2$ corresponds to the maximum velocity of the ionized gas, with $\Delta V$ being the velocity difference in km/s between the narrow and broad kinematic components. The median values of these parameters for the different populations in the BPT-N diagram can be found in Table~\ref{table:summary_properties}. The data presented in Figure~\ref{plot:plot_3x3_kinematics} shows that in most of the cases the broad components are blue-shifted (negative $\Delta V$) with respect to the systemic emission, providing clear evidence that they are outflows rather than just turbulence. The red-shifted component is generally obscured by the host galaxy.

\begin{figure*}
\begin{center}
\includegraphics[width=0.98\textwidth]{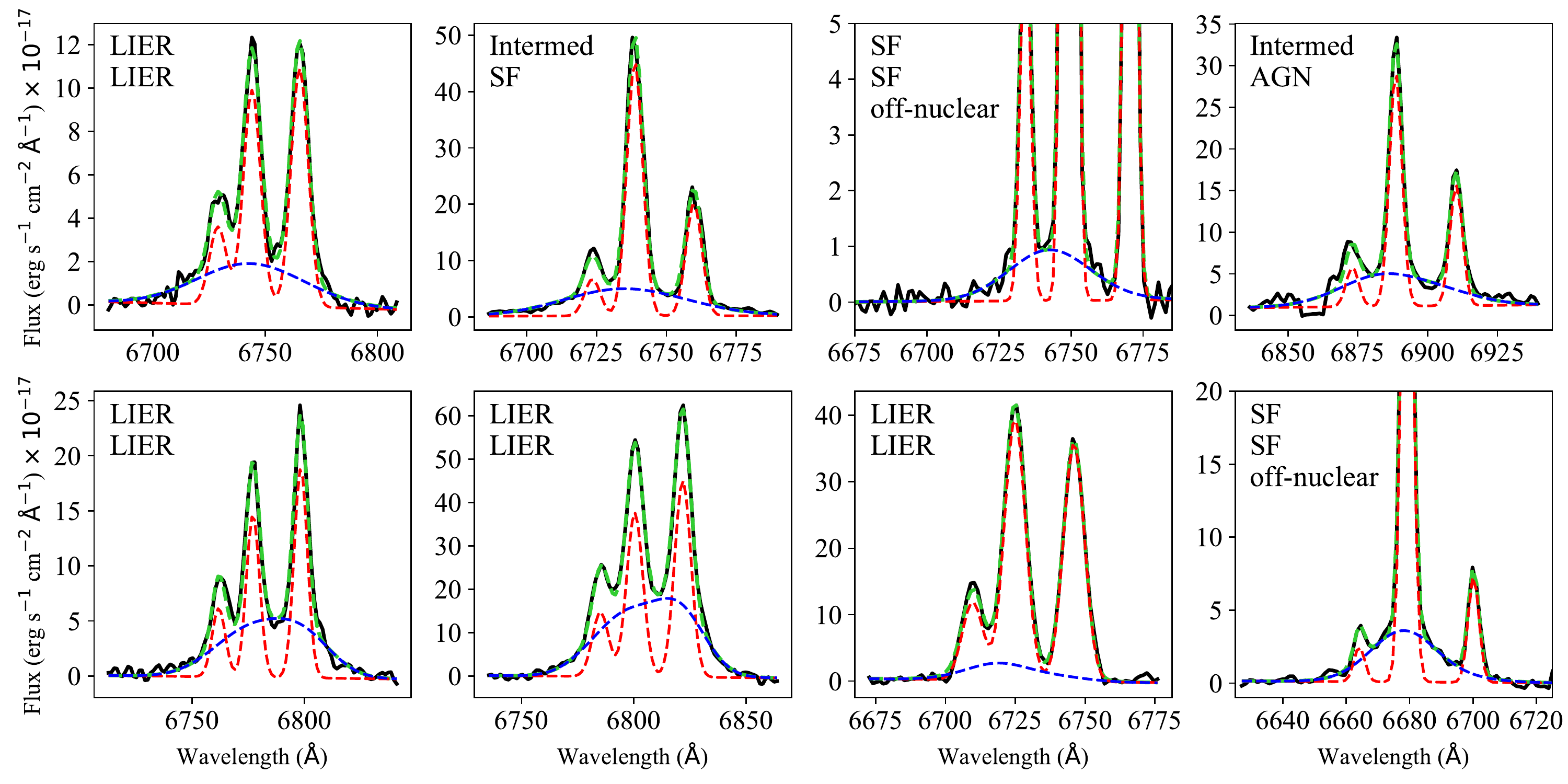}
\caption[sigmavsfr]{Spectral fit to the outflows with the highest $FWHM_{\rm broad}$ in our sample. In each panel we also include the BPT classification based on the BPT-N ($top$) and BPT-S ($bottom$) diagnostic diagrams. We also highlight the two cases of off-nuclear outflows. As in Figure~\ref{plot_2compfit}, the \emph{solid, black} line is the observed spectrum, the \emph{dashed-red} and \emph{dashed-blue} lines correspond to the narrow and broad kinematic components, respectively, and the \emph{dashed-green} line is the sum of these two components. The values of the parameters for the model represented here correspond to the median of their posterior probability density distributions (see Section~\ref{section:spectral_fitting}).}
\label{plot:plot_8highest_fwhm}
\end{center}
\end{figure*}

As shown in the bottom panels of this figure, outflows in LIERs, with a median value of $\sim500$ km/s, have generally associated the largest maximum velocities. In the case of outflows originated in star-forming regions, $V_{\rm max}$ is generally below 500~km/s (log$_{10}$(V$_{\rm max}$)$\sim2.7$), with a median value of $\sim220$~km/s. However, in some cases they reach values above~1000~km/s, as it is the case of the off-nuclear regions mentioned above. Such differences with the type of ionization might be associated with low- and high-velocity shocks originated as the gas expands through the ISM \citep{rupke_outflows_2005, ho_sami_2014}. In fact, as we show in Section~\ref{section:shocks}, in a significant fraction of the outflows there is evidence for ionization by shocks. The higher velocity of the gas in AGN than in SF outflows has been widely observed \citep[][and references therein]{arribas_ionized_2014} and is also found in the recent study of outflows in galaxies at redshifts $0.6-2.7$ by \citet{schreiber_kmos^3d_2018}. Contrary to what is found in this latter work, we do not find a correlation between the $V_{\rm max}$ and stellar mass for AGN sources, although the low number of AGN with detected outflows (12) prevents us from placing strong constraints. 

\begin{figure*}
\begin{center}
\includegraphics[width=0.98\textwidth]{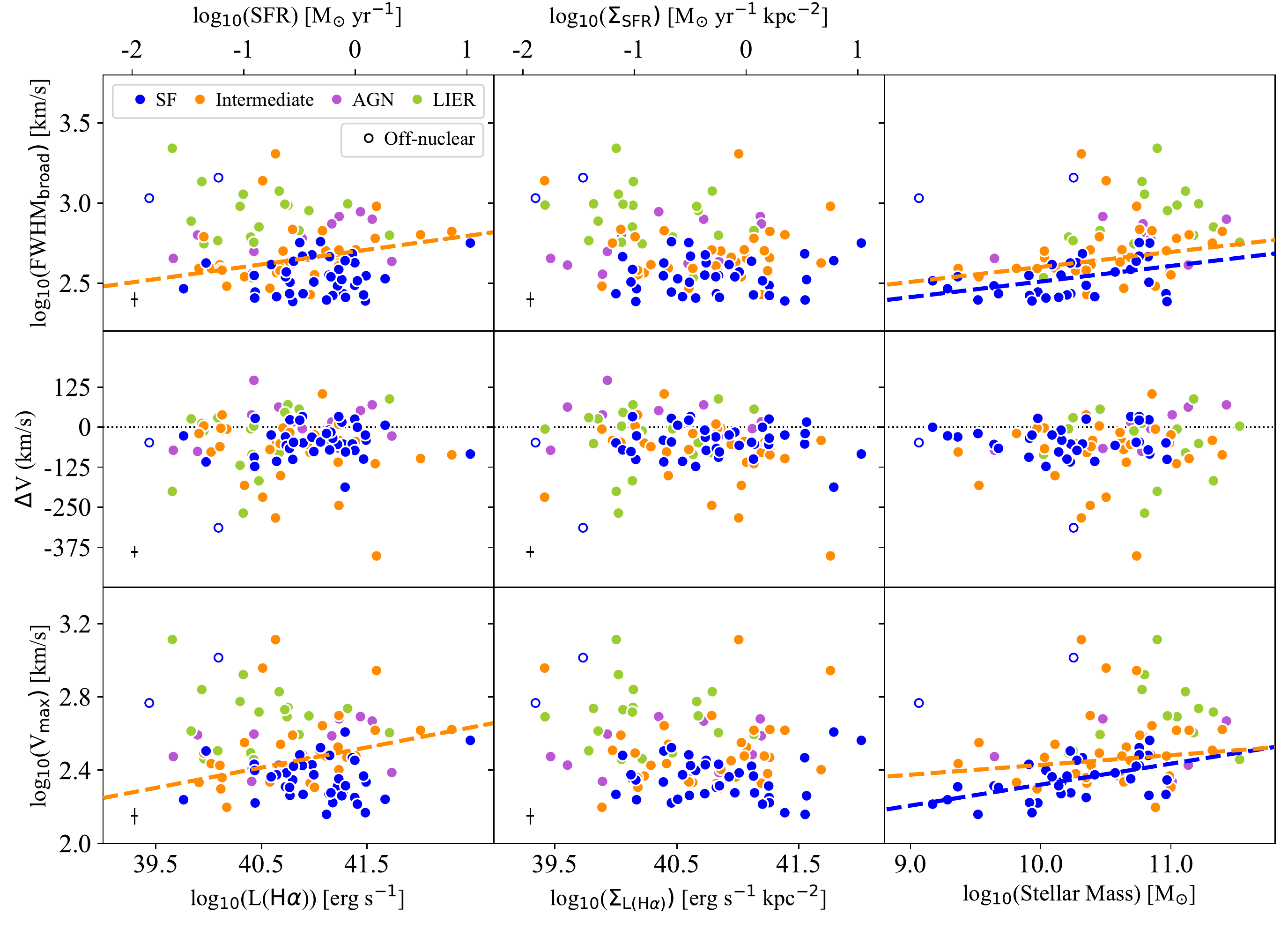}
\caption[sigmavsfr]{Kinematic properties of the ionized gas as a function of $L$(H$\alpha$), $\Sigma_{L({\rm H}\alpha)}$ and stellar mass of the host galaxy for the sources classified according to the BPT-N diagram, using the same colour code as in Figure \ref{plot:bptplots}. In the top axis we show the corresponding $SFR$ and $\Sigma_{SFR}$ values which applies for the SF population only. Empty symbols correspond to off-nuclear regions. We represent $FWHM_{\rm broad}$, the difference in velocity between narrow and broad kinematic components, $\Delta V$, and the maximum velocity of the ionized gas in the outflow, $V_{\rm max}$, estimated as $V_{\rm max} =  |\Delta V| + FWHM_{\rm broad}/2$. The dotted lines at $\Delta V = 0$ in the middle panels are used to separate blue-shifted ($\Delta V < 0$) and red-shifted outflows ($\Delta V > 0$ ). When statistically significant correlations between the different parameters are found for a given population, we show them as thick, dashed lines with the colour corresponding to that population. As shown in the plots we find correlations for the star-forming (blue) and `intermediate' (orange) populations.} 
Typical (mean) error values are shown in the bottom corner of the panels.
\label{plot:plot_3x3_kinematics}
\end{center}
\end{figure*}

In the top axis of Figure~\ref{plot:plot_3x3_kinematics} we also show the $SFR$ and $SFR$ density, $\Sigma_{SFR}$, which applies only for the outflows originated in star-forming regions. We estimate total $SFRs$ following \citet{kennicutt_past_1994} and using the \Ha fluxes from the narrow and broad kinematic components. We apply a dust attenuation correction using the reddening curve from \citet{calzetti_dust_2000} and a colour excess $E(B-V)$ given by the Balmer decrement \citep{dominguez_dust_2013}. In this calculation, a value of \Ha/\Hb~=~2.86 is assumed, which corresponds to a temperature $T=10^4$K and an electron density $N_{\rm e}~=~10^2$cm$^{-3}$ for Case~B recombination, typical values for star-forming regions. Whenever the outflow is only detected in the \Ha line, the \Hb flux used to correct for dust attenuation in the broad component is taken from the single component fit to the
spectra. Again, given the limited spatial resolution of our data, the values of $\Sigma_{SFR}$ estimated here should be considered lower limits. 

We explore now the existence of correlations between the kinematics, star formation properties and stellar mass of the outflows originated in SF regions. To do that we use a Spearman's correlation test to study the distribution of the different parameters, not including the two outliers with $V_{\rm max} >$~500~km/s to reduce the scatter. From this search we find significant a strong correlation ($p$-$value < 0.1$) of stellar mass with respect to $V_{\rm max}$ and a slightly weaker correlation ($p$-$value < 0.3$) with respect to $FWHM_{\rm broad}$. We show these relations in the top- and bottom-right panels of Figure~\ref{plot:plot_3x3_kinematics} as dashed, blue lines. This finding indicates that the kinematics of ionized outflows, traced by $FWHM_{\rm broad}$ and $V_{\rm max}$, increase slightly towards larger stellar masses of the host galaxies, with $log-log$ slopes of $\sim0.39$ and $\sim0.52$, respectively. We find a large scatter as a function of $SFR$ and $\Sigma_{SFR}$ but without any significant trend.

The lack of correlations between outflow kinematics and star formation properties is in contrast with the findings of positive correlations in other works such as the one on local Ultra Luminous Infrared Galaxies (U/LIRGs) by \citet{arribas_ionized_2014}. However, such differences might be due to the lower regime of $SFR$ probed in this work, barely reaching log$_{10}$($SFR$)~$>$1~M$_{\odot}$yr$^{-1}$. In fact, our findings agree with the flattening of the relation between outflow velocity and $SFR$ below log$_{10}$($SFR$)~$\sim1$ M$_{\odot}$yr$^{-1}$ reported in the SDSS stacking analysis of \citet{cicone_outflows_2016}. Considering also neutral outflows, the weak correlation we find between $V_{\rm max}$ and $SFR$ is also in relative agreement with the large scatter and little correlation found by \citet{rupke_outflows_2005} in their study of neutral outflows in local starburst galaxies with $SFRs$ $\sim50-200$ M$_{\odot}$yr$^{-1}$. In fact, they only find a relation when the dwarf galaxies from \citet{schwartz_keck/hires_2004}, whose $SFRs$ are much lower ($\leq0.1$ M$_{\odot}$yr$^{-1}$), are also included. However, such weak correlations are in contrast with those reported by \citet{heckman_implications_2016} for extreme neutral outflows, where a clear correlation is found. Relations between outflow kinematics and stellar mass are more commonly found, both in ionized \citep{cicone_outflows_2016} and in neutral outflows \citep{heckman_implications_2016}. 

 Considering also other type of sources, we also find relations for the `intermediate' population (also removing the objects with $V_{\rm max} >$~500~km/s) between the kinematics, ($V_{\rm max}$ and $FWHM_{\rm broad}$), and both $L$({\rm H}$\alpha$) and stellar mass. The existence of these correlations as a function of stellar mass is somehow expected because part of the ionizing radiation of these objects must come from star formation, which already correlates with stellar mass. However, the correlations with respect to $L$({\rm H}$\alpha$), which is not significant for pure, star-forming regions, indicate that additional sources of ionization in these objects might itself correlate with $L$({\rm H}$\alpha$). LIERs and AGN correspond to more massive systems (on average), but the range in stellar mass (and the limited number of points) is not enough to define a clear correlation.

\subsubsection{Ionization by shocks}
\label{section:shocks}

Many different works on galactic outflows have invoked the presence of shocks in order to reproduce the type of ionization and kinematics observed in the ionized gas \citep[i.e.,][]{monreal-ibero_liner-like_2006,monreal-ibero_vlt-vimos_2010, rich_ngc_2010,westmoquette_spatially_2011, ho_sami_2014, wood_supernova-driven_2015, lopez-coba_star_2017}. These shocks are produced when bubbles of hot gas, originated from supernovae explosions and stellar winds, propagate through the inter-stellar medium (ISM), sweeping up the cool and dense gas. The presence of shock-ionized gas in SF and LIER regions can be identified by the enhancement of the \SII/\Ha ratio, which correlates with the velocity of the gas in shock models \citep{monreal-ibero_vlt-vimos_2010, ho_sami_2014, lopez-coba_star_2017}.  Since the high \SII/\Ha ratios in AGN are not necessarily due to shocks, we do not include them in this analysis.


In Figure \ref{plot:shocks} we explore the presence of ionization from shocks in SF, `intermediate' and LIERs regions classified using the BPT-N diagnostic diagram, showing the velocity dispersion of the gas as a function of the \SII/\Ha ratio, for the narrow (systemic) and broad kinematic components. In this figure we only include values for the spectra where the outflow is detected in all emission lines, as explained in Section~\ref{section:sample_selection}. For reference, we also show in this figure a vertical line at log$_{10}([\mathrm{S}\textsc{ii}]/\mathrm{H}\alpha)>-0.5$, the approximated value at which shock models are able to reproduce the observations \citep{dopita_new_2013, ho_sami_2014}. The data presented here shows a clear trend between the velocity dispersion of the gas in both kinematic components and \SII/\Ha, which can be specially identified in the star-forming population. In general, the \SII/\Ha ratio in the broad kinematic component is larger than in the narrow component and in most of the cases above $-0.5$. This finding indicates that for roughly half of the sample, despite the outflow originates in star-forming regions, the ionization of the outflowing gas could also be due to shocks likely produced when the hot gas propagates through the ISM. 






\begin{figure}
\begin{center}
\includegraphics[width=0.49\textwidth]{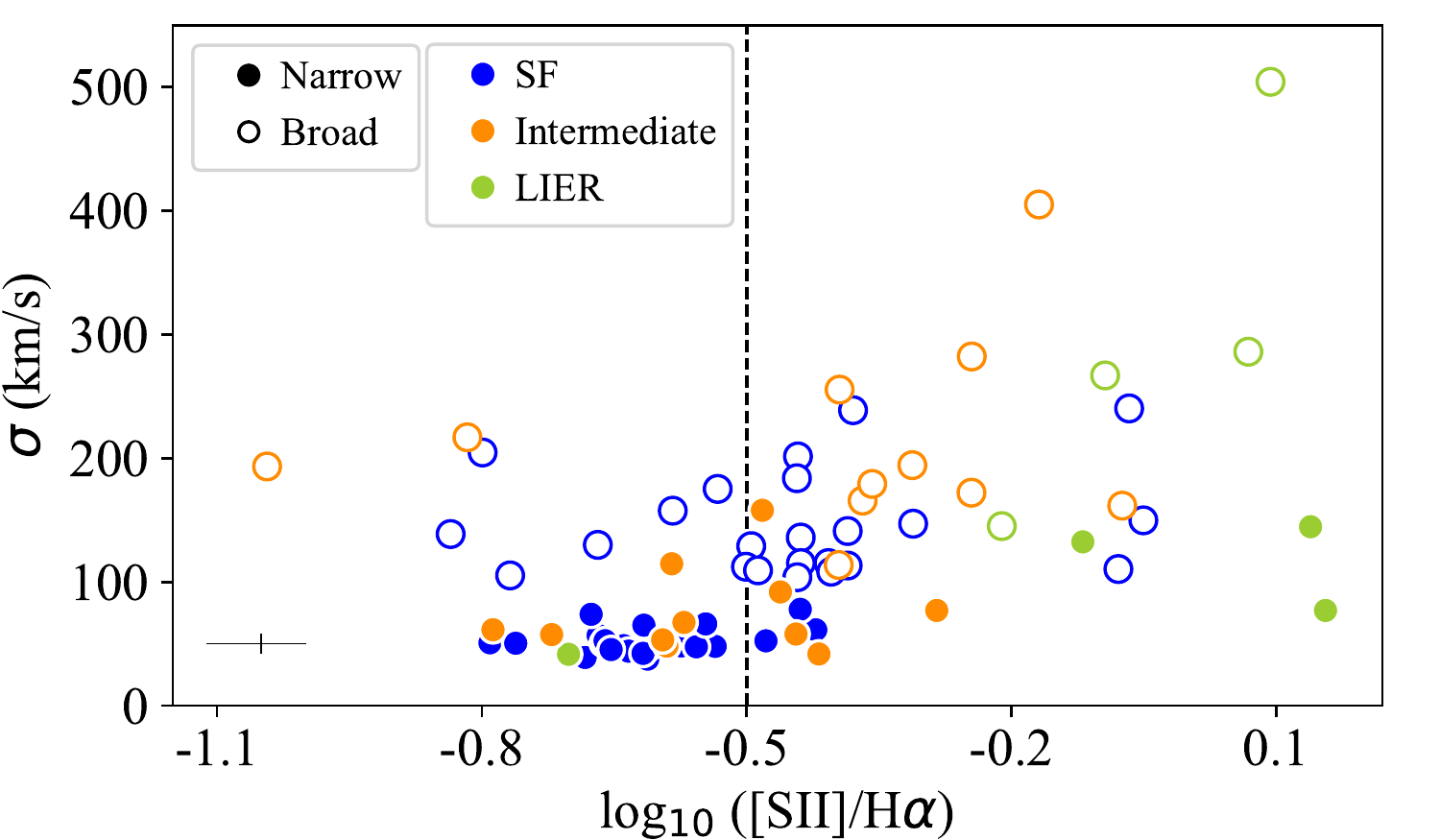}
\caption[sigmavsfr]{Velocity dispersion of the narrow and broad kinematic components as a function of the ratio \SII/\Ha for the sample of outflows in SF regions and LIERs detected in all the emission lines. The vertical line indicates the approximate value at which shock models predict that the gas is ionized by shocks \citep{dopita_new_2013}. The plot shows that there is a correlation between the velocity dispersion of the broad kinematic component and the ratio \SII/\Ha. This ratio is also generally higher than in the narrow one. Typical (mean) error values are shown in the bottom-left corner of the figure.}
\label{plot:shocks}
\end{center}
\end{figure}

\subsubsection{Extinction}
\label{section:extinction}

	The study of the effects of dust extinction provides relevant information about the material that is being entrained and processed by the outflow as well has the effect that it might have on the observability of the spectral signature of ionized outflows in our spectra. As shown in Section~\ref{section:sample_selection}, more than half of the outflows detected in the \Ha line were not detected in other lines such as \Hb, \OIIIb and \SII$\lambda\lambda6717,6731$. This lack of detections could be a consequence of the lower signal-to-noise in these emission lines, partially due to the effects of dust extinction. To explore these effects we estimate the extinction associated to the systemic (narrow) and outflowing (broad) gas components, $A_{{\rm V}, {\rm broad}}$ and $A_{{\rm V}, {\rm narrow}}$, respectively, following the same procedure explained in Section~\ref{section:kinematics}. We compare the two extinction values in Figure~\ref{plot:extinction}, using the values corresponding to the spectra classified using the BPT-N diagram (Section \ref{section:bpt}) where the broad kinematic component is detected in the \Ha and \Hb lines (54 objects). In Table \ref{table:summary_properties} we also show the median values for all the sources and for the different populations.

\begin{figure}
\begin{center}
\includegraphics[width=0.45\textwidth]{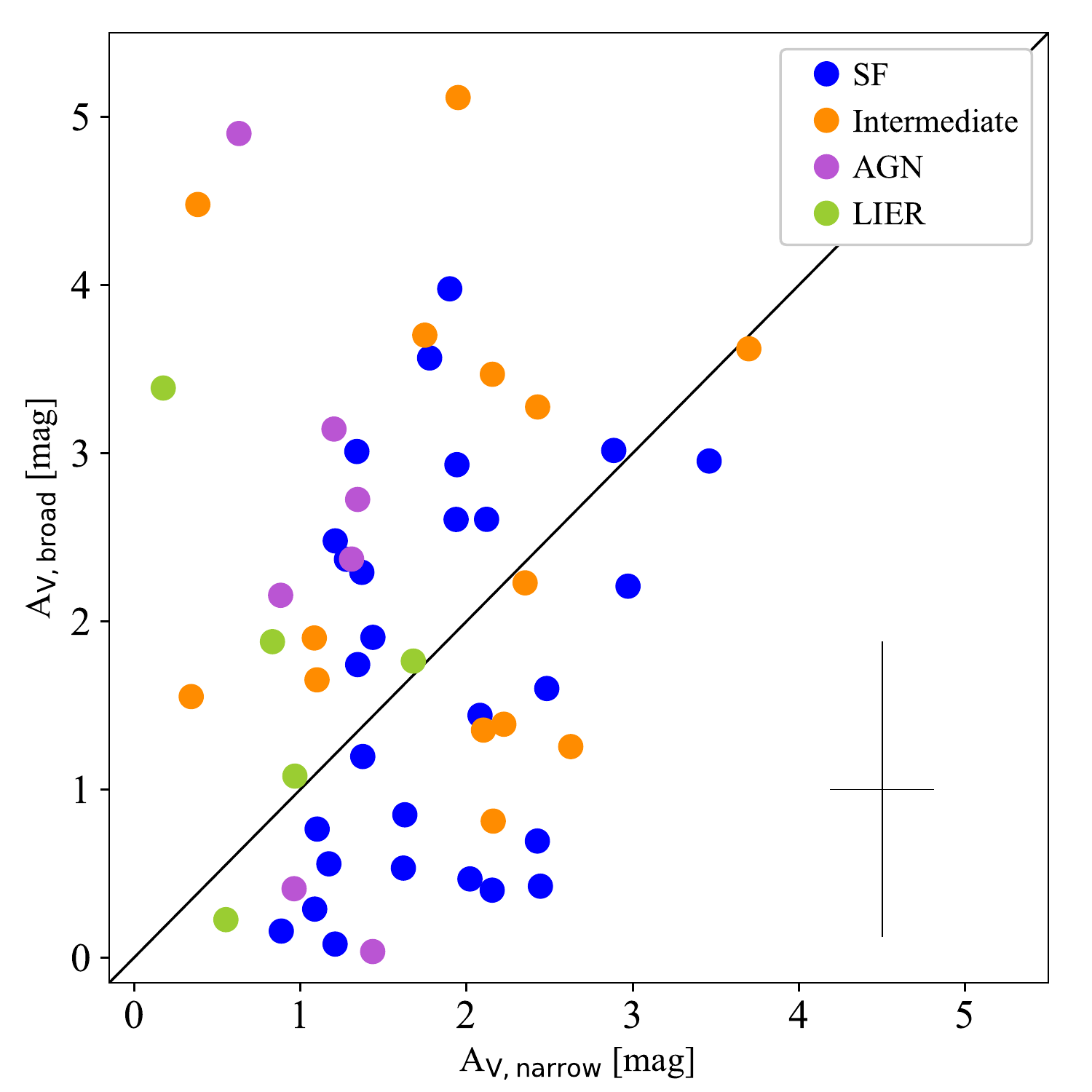}
\caption[sigmavsfr]{Comparison of the extinction, $A_{\rm V}$, for the narrow and broad kinematic components for the sample of outflows detected in the \Ha and \Hb lines (54 objects). The solid line is the one-to-one relation. Typical (mean) error values are shown in the bottom-right corner of the figure.}
\label{plot:extinction}
\end{center}
\end{figure}

The scatter in the values of extinction for both components is quite large, with some values basically consistent with no dust and others reaching values up to $4-5$ magnitudes in $A_{\rm V}$. Considering all the sources, the extinction in the broad component is slightly higher (1.5 and 1.9 for narrow and broad, respectively); however, this difference disappears if only outflows in SF or `intermediate' regions are considered. For AGN and LIERS, although the numbers are low, there seems to be a trend to have higher extinction in the broad component than in the narrow one, a result similar to what is found in type~2~QSOs \citep{villar_martin_triggering_2014}. Despite the differences in $A_{\rm V}$ between the two components are not significant, the effects from extinction can be quite severe, specially in an outflowing component intrinsically broad and with a generally low contribution to the total line flux. Therefore, such high extinction values could contribute to the lack of detections in other spectral lines apart from \Ha in a significant fraction of the spectra, specially at shorter wavelengths where the effects of dust attenuation are more important.


\begin{table*}
\caption[Summary of outflows properties]{Summary of the properties of outflows and host regions for the different ionizing sources identified using the BPT-N diagram. We show the median values and $MAD$ of each parameter in each case.}
\label{table:summary_properties}
\begin{tabular}{llccccc}
\hline
\hline
& & SF & Intermediate & LIER & AGN & All sources\\
\hline
$FWHM_{\rm broad}$ (km/s) &  & 346 (76) & 452 (75) & 896 (238) & 550 (135) & -\\ 
$V_{\rm max}$ (km/s)  & & 225 (43) & 295 (67) & 497 (104) & 346 (90) & -  \\   
F(\Ha)$_{\rm broad}$/F(\Ha)$_{\rm narrow}$ & & 0.20 (0.12) & 0.22 (0.15) & 0.45 (0.33) & 0.49 (0.26) & -\\
\hline
$A_{\rm V}$& narrow & 1.7 (0.4) & 2.1 (0.4)  & 0.8 (0.3)  & 1.2 (0.2) & 1.5 (0.5) \\
& broad & 1.7 (1.0) & 2.1 (1.0) & 1.8 (0.7) & 2.4 (0.8) & 1.9 (1.1)\\
$N_{\rm e}$ (cm$^{-3}$)& narrow & 126 (77) & 248 (81) & 267 (28) & 269 (140) & 222 (122) \\    
& broad & 302 (123) & 281 (150) & 706 (264) & 227 (54)  & 281 (144) \\    

\hline
\end{tabular}
\end{table*}

\subsubsection{Density of the gas}
\label{section:density}

An important parameter that characterizes the properties of the systemic and outflowing components is the electron density of the ionized gas associated to them. The electron density can be estimated using the temperature of the ionized gas and the ratio between the fluxes from the \SII$\lambda\lambda6717,6731$ emission lines \citep{osterbrock_astrophysics_1989}. 
Assuming a typical temperature for the ionized gas of $T=10^4$K and following the expressions derived by \citet{sanders_mosdef_2016}, we calculate the electron densities of the two kinematic gas components for the 45 objects where outflows are detected in all the emission lines. The individual values are shown in Figure~\ref{plot:densities} and the mean values presented in Table~\ref{table:summary_properties}. Despite of the relatively large uncertainties of the individual values derived using this method \citep[e.g.,][]{genzel_sins_2011, villar_martin_triggering_2014,perna_x-ray/sdss_2017}, we can extract some conclusions.  The gas in the outflow tends to be, on average, more dense than the one associated to the narrow component. In particular, in the star-forming regions the median values are 126~cm$^{-3}$ (Median Absolute Deviation $(MAD) = 77$) and 302~cm$^{-3}$ ($MAD=131$) for the narrow and broad components, respectively. 

This finding of overall higher densities in the outflowing components is in agreement, within the uncertainties, with what is found in other works \citep{arribas_ionized_2014, villar_martin_triggering_2014}.  In summary, the large scatter observed in the Figure \ref{plot:densities} demonstrates the wide range of densities that the ionized gas can have. However, the fact that the gas tends to be more dense in the outflowing component might indicate that the pressure is higher, which can also be a consequence of shocks \citep{ho_sami_2014}. 


\begin{figure}
\begin{center}
\includegraphics[width=0.45\textwidth]{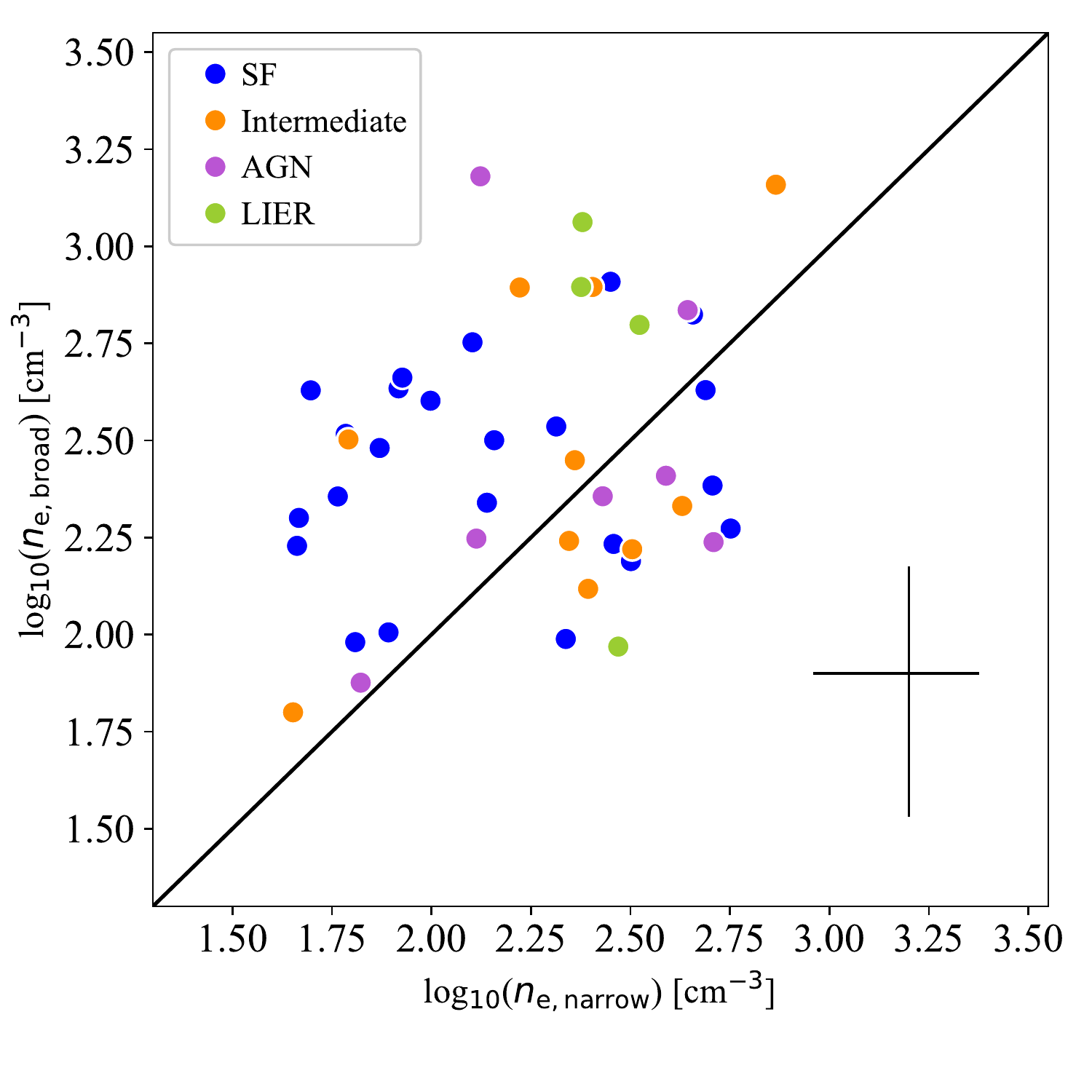}
\caption[plot_densitiesI]{Comparison of the electron densities, $n_{\rm e}$, estimated using the \SII$\lambda6717,6731$ lines, for the narrow and broad components. Typical (mean) error values are shown in the bottom-right corner of the figure. }
\label{plot:densities}
\end{center}
\end{figure}


\subsection{Global effects of the outflows on the host}
\subsubsection{Upper estimates of the mass-loading factor}

In this section we explore what is the effect that the ionized outflows studied here can have in the star formation activity of the regions where they are detected. The standard method to evaluate the impact of outflows is to compare the amount of gas that is being ejected by the outflow, referred to as the outflowing mass rate, $\dot M$, with the amount of gas that is being converted to stars, given by the SFR. The ratio between these two quantities is known as the mass-loading factor, $\eta = \dot M/SFR$. To estimate $\eta$
we follow Equation 4 in \citet{arribas_ionized_2014}, assuming an outflow extension of 0.7~kpc, which is the typical extension found in the extended studies of \citet{bellocchi_vlt/vimos_2013}. As mentioned in \citet{arribas_ionized_2014}, this derivation of the mass-loading factor assumes that all the gas moves at maximum velocity and does not account for the intrinsic structure of the outflow, which can lead to differences by factors of $2-3$ in the estimations of $\eta$. Moreover, its estimation is also subject to the large uncertainties in the values of the electron densities (see Section \ref{section:density}). Based on these assumptions and uncertainties, our estimates of the mass-loading factor should not be considered at face value but instead as upper limits. Although higher resolution studies can obviously characterise better the outflow structure and geometry and therefore provide more accurate estimates of $\eta$, the present MANGA data allow us to constrain whether the outflows we study are expected to have or not a significant impact on their surroundings.

\begin{figure}
\begin{center}
\includegraphics[width=0.49\textwidth]{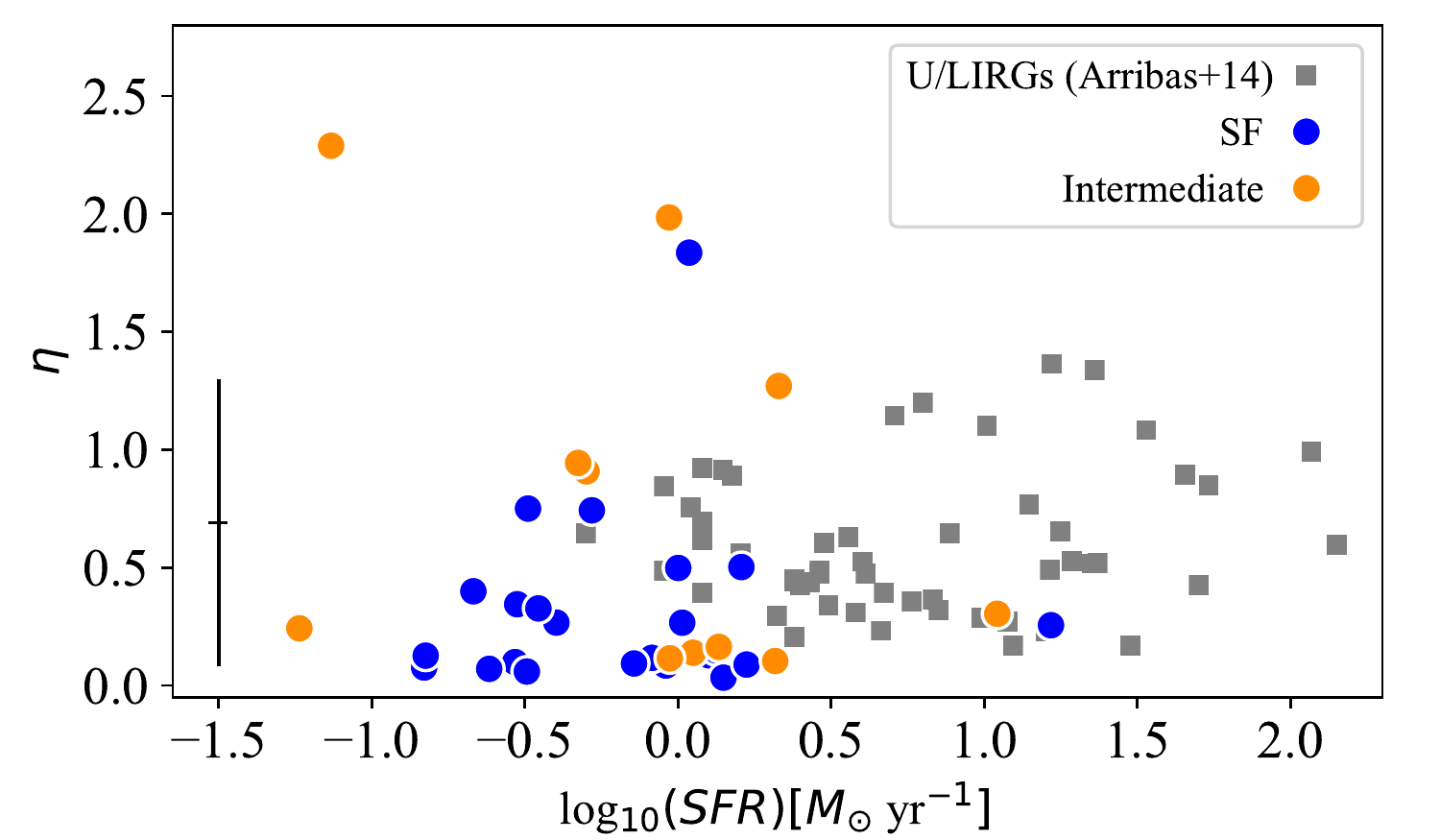}
\caption[masloading]{Mass-loading factor, $\eta$, as a function of $SFR$ for the sample of ionized outflows in SF regions detected in all the emission lines we explore. For comparison, we also show the values for the integrated properties of U/LIRGs from \citet{arribas_ionized_2014}. Typical (mean) error values are shown in the bottom-left corner of the figure.}
\label{plot:massloading}
\end{center}
\end{figure}

In Figure \ref{plot:massloading} we show the values of the mass-loading factor, $\eta$, for the sample of outflows in SF and `intermediate' regions detected in all the emission lines studied here, as a function of the SFR. We also include the `intermediate'  population in this plot because part of the ionization in these sources must come from star formation. The values of $\eta$ we obtain are in general quite low, with a median value $\eta\sim0.25$ considering SF and `intermediate' objects; only a few outflows have values larger than 1. Considering that only a small fraction of the gas will be moving at the maximum velocity, our findings indicate that the outflows studied here lead to little, if any, suppression of star formation. In the figure we also include the data corresponding to the integrated values obtained for U/LIRGs by \citet{arribas_ionized_2014} which, despite covering a higher range of $SFRs$ (partially due to be integrated data), generally have larger mass-loading factors than those measured here. Comparing also with the study of outflows at $z=0.6-2.7$ by \citet{schreiber_kmos^3d_2018}, they find slightly lower values ($\eta\sim0.1-0.2$) than our median estimate, although the SFRs they probe are significantly higher (log$_{10}(SFR)\sim[-1, 2.5]$). This similarity could be an indication that, at least since $z=2.7$, the impact of ionized outflows driven by star formation in the star formation activity has not been significant. We note here that the estimation of the mass-loading factor only considers the gas that is in the ionized phase and that this value might vary significantly when the neutral and molecular phases are also accounted for. 

Finally, as shown in Table \ref{table:summary_properties}, the ratio between the \Ha fluxes in the narrow and broad components, F(\Ha)$_{\rm broad}$/F(\Ha)$_{\rm narrow}$, increases significantly from SF and `intermediate' with roughly 20\%, to AGN and LIERs, where the broad components encompasses around half of the total flux in the narrow ones. This result indicates that outflows in AGN and LIERs entrain relatively larger amounts of ionized gas than those originated in SF and `intermediate' regions.

\begin{figure}
\begin{center}
\includegraphics[width=0.49\textwidth]{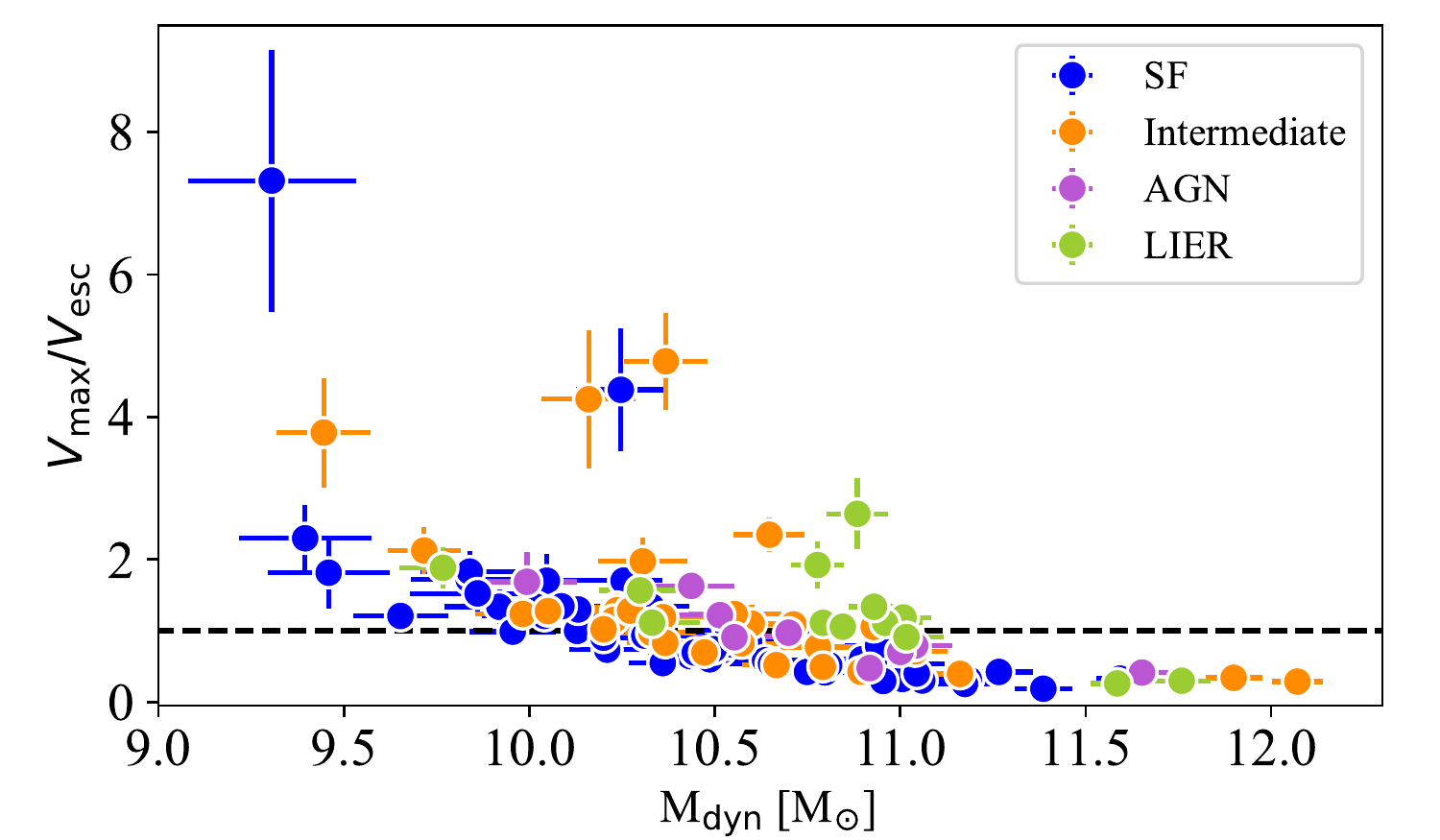}
\caption[Vmax/Vesc vs dynamical mass.]{Ratio between the maximum velocity of the gas, $V_{\rm max}$, and the escape velocity, $V_{\rm esc}$ as a function of the dynamical mass of the galaxies. The horizontal line corresponds to $V_{\rm esc}/V_{\rm max} = 1$.}
\label{plot:vmaxvesc_vs_mdyn}
\end{center}
\end{figure}

\subsubsection{Gas kinematics and escape velocity}

The metal content of galaxies has been proven to be a property that is strongly correlated with the mass of the galaxies, in a way that more massive galaxies have higher metallicities than lower-mass ones, following the well-defined mass-metallicity relation \citep{tremonti_origin_2004, mannucci_fundamental_2010}. This relation is thought to be held, at least partially, due to the action of outflows regulating the amount of metals formed and kept inside the galaxy \citep{chisholm_metal-enriched_2018}. They can regulate the formation of metals by suppressing the star formation and/or drive metal-rich gas out of the galaxies to the IGM \citep{brooks_origin_2007, finlator_origin_2007}. In both cases, for a given outflow power, low-mass galaxies are expected to eject more material due to their shallower potential wells. Moreover, the presence of outflows has also been invoked to explain the high metal content of the IGM at high redshift \citep{aguirre_metal_2001}.

The standard method to evaluate whether the gas entrained by an outflow can be expelled from a galaxy is to compare the maximum velocity of the outflowing gas with the escape velocity of the galaxy. This escape velocity can be estimated from the circular velocity of the gas assuming a singular truncated isothermal sphere, as in \citet{heckman_absorption-line_2000}. Given that we have not estimated the circular velocity, we follow the method described in \citet{arribas_ionized_2014} to estimate escape velocities using the dynamical masses of the galaxies. The dynamical masses are derived from equation 3 in \citet{bellocchi_vlt/vimos_2013}, using the effective radius, the rotational velocity and the velocity dispersion. In our case, we benefit from the recent release of the MaNGA Value Added Catalogue, which contains a large number of derived parameters for MaNGA DR2 galaxies \citep{sanchez_pipe3d_2016, sanchez_ssdss_2017}. From this catalogue\footnote{\url{https://data.sdss.org/sas/dr14/manga/spectro/pipe3d/v2\_1\_2/2.1.2/manga.Pipe3D-v2\_1\_2.fits}} we estimate the dynamical masses of the galaxies using the effective radius ($R_{eff}$), the stellar velocity in the central 2.5 arcsec ($\sigma_{\rm cen}$) and the velocity/dispersion ratio ($v/\sigma$) for the stellar populations within 1.5 $R_{eff}$. We note here that $v/\sigma$ and $\sigma_{\rm cen}$ are estimated using different areas, which could introduce a bias when the velocity dispersion varies significantly within 1.5 $R_{eff}$. However, we expect this difference to be within a factor of $\sim2$ based on the radial velocity dispersion profiles across different Hubble types \citep{falcon-barroso_stellar_2017}. 

In Figure \ref{plot:vmaxvesc_vs_mdyn} we show the ratio between  $V_{\rm max}$/$V_{\rm esc}$ as a function of the dynamical mass $M_{dyn}$ of the galaxies, for our sample of outflows, colour-coded following the classification of the BPT-N diagram (Section \ref{section:bpt}). Here, as in \citet{arribas_ionized_2014}, we have estimated the escape velocity at 3~kpc for an isothermal sphere truncated at 30~kpc\footnote{The choice of the truncated value of the sphere at 30 kpc is somewhat arbitrary. However, we note that the values of $V_{\rm esc}$ change by less than 20\% if the truncated value is placed at 100 kpc, introducing no significant changes in our results.}. Despite all the uncertainties associated to the estimation of $V_{\rm max}$ and $V_{\rm esc}$, we find a clear trend between their ratio and the dynamical mass of the galaxies. At low dynamical masses the velocity of the ionized gas in the outflow is high enough to overcome the gravitational pull of their hosts, whereas at larger dynamical masses, the increase of the escape velocity reduces the ratio $V_{\rm max}$/$V_{\rm esc}$, favouring the withholding of the gas entrained by the outflows. This result is similar to what is found in the study of U/LIRGs by \citet{arribas_ionized_2014}, although with our sample we extend this trend towards lower dynamical masses. This result agrees nicely with the steep increase of gas-phase metallicity with stellar mass up to $10^{10.5}$ M$_{\odot}$ found by \citet{tremonti_origin_2004}. These results indicate that there is a clear connection between the escape velocity and the metallicity of the galaxies, as also reported recently by \citet{barrera-ballesteros_sdss-iv_2018}. 


In summary, our results indicate that ejection of gas entrained by ionized outflows to the IGM is more relevant at low dynamical masses, reducing its efficiency towards larger masses. Although the outflow velocities are higher for massive galaxies (see Figure~\ref{plot:plot_3x3_kinematics}), in general they are not high enough to abandon the potential well of their hosts, retaining the gas and maintaining their high gas metallicities.

\subsubsection{In-situ star formation in ionized outflows}
\label{section:insituSF}
Recent works have reported, first in a single system \citep{maiolino_star_2017} and more recently in several objects \citet{gallagher_widespread_2018}, the detection of star formation taking place inside galactic outflows, this latter work also using data from MaNGA DR2. This result, predicted by some theoretical models \citep{ishibashi_can_2013, zubovas_galaxy-wide_2014}, is of high relevance because it demonstrates that feedback in the form of outflows can enhance star formation, providing a new route for producing stars. Moreover, as explained in \citet{gallagher_widespread_2018}, the occurrence of star formation inside outflows could help explaining current paradigms such as the early formation of spheroids, the establishment of the relation between black holes and their hosts and other observed properties. 

	The identification of outflowing gas ionized by stars does not by itself prove the presence of ionizing starts within the outflow. It could also be be possible that the photons from stars located in the disc ionize the outflowing gas. To distinguish between these two scenarios, and following \citep{maiolino_star_2017} and \citet{gallagher_widespread_2018}, here we explore the values of the ionization parameter, which is expected to be significantly lower when the gas is far from the ionizing stars (all the other properties being equal). We do that by comparing the values of the ionizing parameter, $U$, which describes the degree of ionization of the gas, for both kinematic components. Following \citet{diaz_chemical_2000}, we calculate this parameter as the ratio between \OIIIb and the sum of the {$[\mathrm{O}\textsc{ii}]\lambda\lambda3727,3730$} lines\footnote{For this analysis we repeat the same fitting procedure explained in Section \ref{section:spectral_fitting} but including in the fits the {$[\mathrm{O}\textsc{ii}]\lambda\lambda3727,3730$} lines. In this case the subtraction of the stellar component is done using the MILES library \citep{sanchez-blazquez_medium-resolution_2006}, whose wavelength coverage is 3525-7500\AA. To select the sample we apply the same selection criteria as in Section \ref{section:sample_selection} to the {$[\mathrm{O}\textsc{ii}]\lambda\lambda3727,3730$} lines.}. Here we use only regions hosting outflows ionized by star formation, for which we select the sources whose broad component is located in the SF region of the BPT-N diagram in the bottom panels of Figure~\ref{plot:bptplots}. We note that this way of estimating $U$ is subject to differences in the density of the gas (favouring larger values of $U$ for higher densities) and to the effects of dust extinction (favouring higher values of $U$ if the extinction is high). Note, however, that the electron densities for the narrow and broad components in star-forming galaxies are consistent with each other within a factor of $\sim1.5$ (Table~\ref{table:summary_properties}) and that we have estimated $A_{\rm V}$ for each component, therefore our $U$ estimates account for them. Following \citet{gallagher_widespread_2018} we also estimate the parameter $R_{23}$, defined as the ratio {($[\mathrm{O}\textsc{iii}]\lambda\lambda4960,5007$ + $[\mathrm{O}\textsc{ii}]\lambda3727$)/\Hb}, for both components. 

The results, presented in Figure~\ref{plot:ionization_parameter}, show that the ionizing parameter $U$ (traced by the ratio \OIIIb/$[\mathrm{O}\textsc{ii}]\lambda\lambda3727,3730$) is quite similar in both kinematic components, with only one case where the ratio $U_{\rm narrow}$/$U_{\rm broad}$ is larger than $\sim10$. Given that $U$ decreases with the square of the distance, we can use this result to evaluate whether the star formation ionizing the outflowing gas is coming from the disk of the galaxy or from inside the outflow. To do that, since we do not have accurate estimates of the sizes of the star-forming regions nor the extension of the outflows, we use standard values from the literature: typical sizes of star-forming knots are $20-100$~pc \citep[e.g., ][]{miralles-caballero_extranuclear_2012} whereas the extension of outflows in similar galaxies is $\sim0.7$~kpc \citep[][median value from Table 5]{bellocchi_vlt/vimos_2013}. Assuming these sizes, if the stars ionizing the outflowing gas resided in the disk of the galaxies (at a distance typically $\sim7\times$ the size of the region), the ionizing parameter $U$ in the broad component would be a factor $\sim50$ lower than observed. However, the values shown in Figure~\ref{plot:ionization_parameter} do not suggest such large differences, indicating that our results are consistent with star formation taking place within the ionized outflows, as previously reported by \citet{maiolino_star_2017} and \citet{gallagher_widespread_2018}. We note here that this result only holds under the assumption (based on estimates from other works) that the outflowing regions are significantly larger than individual star-forming ones.

Finally, given that we only detect outflows in $\sim2\%$ of the \Ha-emitting regions and that the upper estimates of the mass-loading factors are generally low ($\eta\sim0.25$; Section~\ref{plot:massloading}), our results indicate that any in-situ star formation in the galaxies studied here would not have a strong impact in their evolution. 


\begin{figure}
\begin{center}
\includegraphics[width=0.49\textwidth]{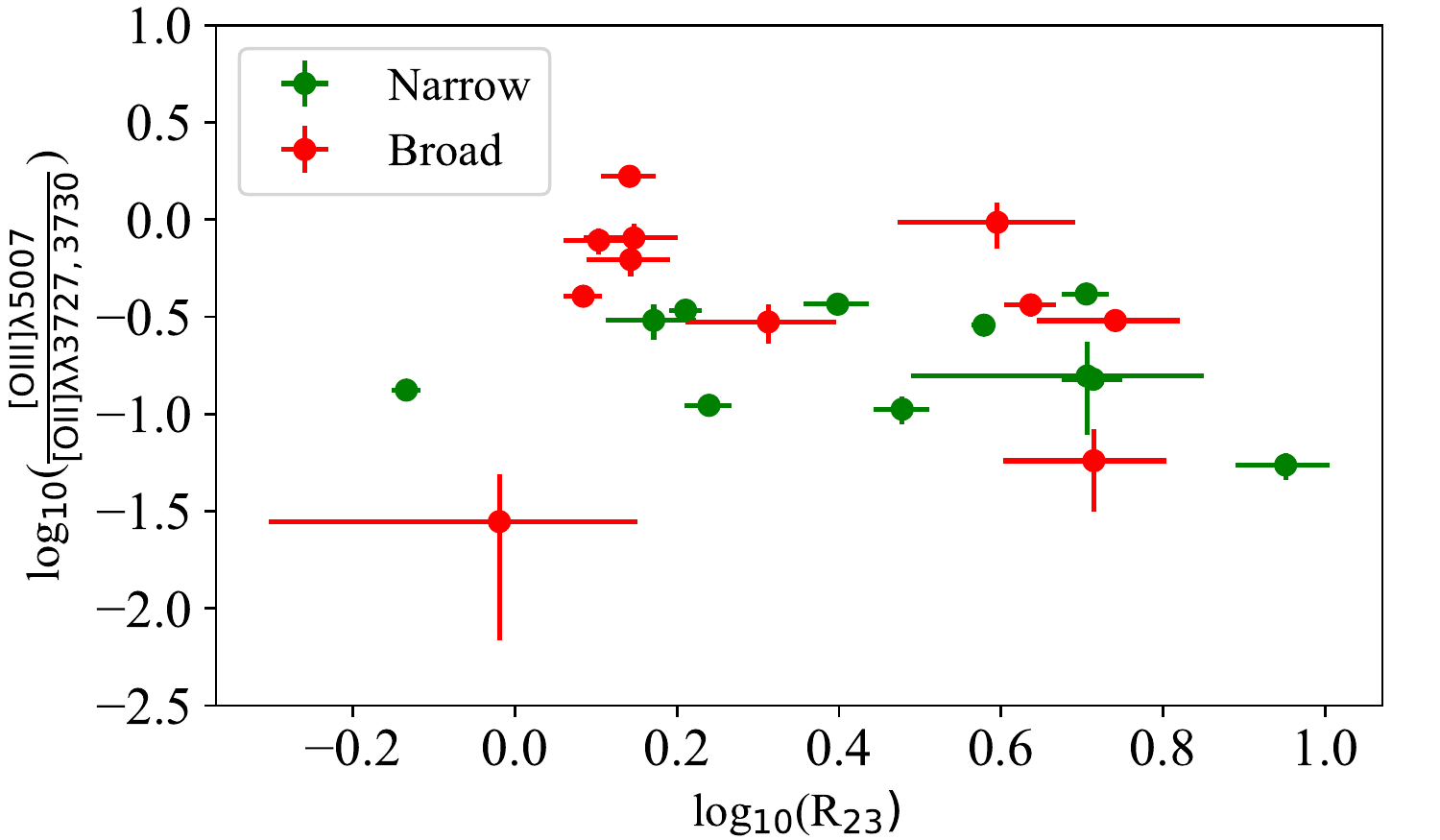}
\caption[fit to SII]{Comparison of the ionizing parameter, $U$, traced by the ratio \OIIIb/$[\mathrm{O}\textsc{ii}]\lambda\lambda3727,3730$, as a function of the $R_{23}$ parameter (see text for details) for both kinematic components, similar to Figure 7 in \citet{gallagher_widespread_2018}. Values take into account individual estimates for $A_{\rm V}$. Here we show only the cases where the broad component is ionized by star formation. Since the ionizing parameter is similar in both components, the outflowing gas must be ionized by in-situ star formation (see Section \ref{section:insituSF} for details).}
\label{plot:ionization_parameter}
\end{center}
\end{figure}

\section{Summary and Conclusions}
\label{sec:summary}

In this work we have presented the results from a systematic search and characterization of ionized outflows in nuclear and off-nuclear \Ha-emitting regions of more than 2700 nearby galaxies ($z\leq0.15$) from MaNGA DR2. The galaxies have stellar masses in the range $\sim10^{9}-10^{12}$~M$_{\odot}$ and the $SFRs$ span from $\sim0.01$~to~$\sim10$~M$_{\odot}$yr$^{-1}$. In total we have identified and analyzed the spectra of $\sim5200$ \Ha-emitting regions from $\sim1500$ individual systems, with a mean of $\sim3.5$ regions per galaxy, searching for spectral signatures of ionized outflows. Ionized outflows are detected, at least in the \Ha line, in a total of 105 individual \Ha-emitting regions, corresponding to 103 individual galaxies. From this sample, in 45 cases the secondary component is also detected in \Hb, \OIIIb and \SII$\lambda6717,6731$. In this work we extend the study of ionized outflows to regions with $SFRs$ as low as $\sim0.01$~M$_{\odot}$yr$^{-1}$, a $SFR$ regime lower than those explored by previous works on individual systems.

We have explored the differences in the kinematics and properties of the ionized gas for the systemic and outflowing components in regions classified as star-forming, `intermediate', LIER and AGN based on the type of ionization, and analysed the feedback effects. The main results we have obtained in our work can be summarized as:

\begin{itemize}
\item  
We detect ionized outflows in $\sim2$\% of the \Ha-emitting regions and in $\sim7$\% of the \Ha-emitting galaxies studied in this work. The detection rate of ionized outflows increases with \Ha flux and stellar mass of the host galaxies. Considering all the \Ha-emitting regions identified in this work, their incidence varies from less than $2\%$ at fluxes below $\sim1.0\times10^{-14}$~erg~s$^{-1}$cm$^{-2}$ to more than $8\%$ in regions with fluxes above that value. Regarding mass, their incidence increases at $\sim10^{10}$M$_{\odot}$, when they are detected in $\sim3\%$ of the cases.\\

\item  
We find only two off-nuclear outflows, located at distances larger than 1~kpc from their galaxies' centers. These outflows show extreme gas kinematics ($V_{\rm max} > 500$km/s and $FWHM_{\rm broad} > 1000$~km/s) although they are only detected in the \Ha line. The origin of such remarkable features is still unclear and will be explored in more detail in a separate paper (Rodr\'iguez Del Pino et al. in prep).  \\

\item We find significant differences in the gas kinematics of outflows originated in regions with different type of ionization. Regions characterized by LIER emission host the outflows with more extreme kinematics ($FWHM_{broad}\sim$~900~km/s), followed by those originated in AGN, `intermediate' and SF regions ($FWHM_{broad}\sim$~550~km/s, $\sim450~$km/s, $\sim350~$km/s, respectively). A similar result is found for the maximum velocities of the gas, $V_{\rm max}$. These differences might be due to the presence of of low- and high-velocity shocks, given the high incidence of outflowing, shock-ionized gas detected in our sample. \\

\item Considering only pure, SF regions, we find significant correlations between the kinematics of the outflowing gas and the stellar mass of the host galaxies, implying higher $FWHM_{\rm broad}$ and $V_{\rm max}$ with increasing stellar mass ($log-log$ slopes of $\sim0.39$ and $\sim0.52$, respectively). However, no significant correlations are found between gas kinematics and star formation properties $SFR$ and $\Sigma_{SFR}$, probably due to the lower $SFR$ regime probed in this study. Moreover, the gas kinematics in `intermediate' regions  correlate  both with the stellar mass $L$(H$\alpha$), probably as a consequence of the additional mechanisms responsible for the ionization of the gas in these sources. \\

\item Although the uncertainties in the estimation of the electron densities are relatively large, the gas is generally denser in the outflowing component, pointing towards a shock-related origin of the outflows.\\

\item We estimate mass-loading factors, $\eta$, generally below~1, with a median value $\eta\sim0.25$ considering both SF and `intermediate' regions. Therefore, despite all the uncertainties associated to the mass loading factor estimates, the ionized outflows studied in this work do not suggest a strong impact in the star formation activity in the host regions. \\

\item Our analysis show that only at low masses ($M < 10^{10}$M$_{\rm sun}$) part of the gas entrained by the outflows is able to escape the gravitational well of the galaxy. At high masses, outflows are probably only able to produce gas fountains that would return the gas to the galaxies. This relation between the fraction of gas that can abandon the galaxy as a function of mass could help explaining the existent correlation between the gas-phase metallicity and the stellar mass of galaxies \citep{tremonti_origin_2004, mannucci_fundamental_2010}. \\

\item The values of the ionization parameter measured in the systemic and outflowing components in regions hosting outflows ionized by star formation are consistent with in-situ star formation, under the assumption (based on estimates from other works) that the outflowing regions are significantly larger than individual star-forming ones. This result implies that at least a fraction of our sample of outflows are compatible with the recently reported finding of star formation taking place inside ionized outflows \citep{maiolino_star_2017, gallagher_widespread_2018}. However, although the uncertainties are large and we only probe the ionized phase of the outflowing gas, the low mass-loading factors we measure indicate that in-situ star formation does not have a significant impact in the galaxies studied here.\\
\end{itemize}

\section*{Acknowledgements}

We thank the anonymous referee for comments and suggestions that have contributed to improve the paper. BRP, SA and LC acknowledge support from the Spanish Ministry of Economy and Competitiveness through grants ESP2015-8964 and ESP2017-83197. JPL acknowledges support from the Spanish Ministry of Economy and Competitiveness through grant AYA2017-85170-R. MVM acknowledges support from the Spanish Ministry of Economy and Competitiveness through the grant  AYA2015-64346-C2-2-P. Funding for the Sloan Digital Sky Survey IV has been provided by the Alfred P. Sloan Foundation, the U.S. Department of Energy Office of Science, and the Participating Institutions. SDSS-IV acknowledges support and resources from the Center for High-Performance Computing at
the University of Utah. The SDSS web site is www.sdss.org. This research made use of Astropy, a community-developed core Python package for Astronomy \citet{astropy_collaboration_astropy_2018}. This project makes use of the MaNGA-Pipe3D dataproducts. We thank the IA-UNAM MaNGA team for creating this catalogue, and the ConaCyt-180125 project for supporting them. 

\bibliographystyle{mn2e}
\bibliography{refs}

\label{lastpage}

\end{document}